\documentclass[12pt,a4paper,dvips]{article}
\usepackage{a4p}
\usepackage{cite,mcite}
\usepackage{graphicx}
\usepackage{physics }
\usepackage{l3_title,ifthen}

\journalname{Phys. Lett. B}

\date{July 31, 2003}

\preprint{2003-059}

\def\dm        {\ensuremath{\Delta M}}
\def\susy#1{\ensuremath{\tilde{\mathrm{#1}}}}%
\def\slepton   #1{\ensuremath{\susy{\ell}^{#1}}}

\def\neutralino#1{\ensuremath{\susy{\chi}_{#1}^0}}
\def\EV30{\ensuremath{E_{\mathrm{v30}}}}%
\def\Ebar{\ensuremath{E\hspace{-.23cm}/\hspace{+.01cm}}}
\def\EM25{\ensuremath{\Ebar_{25}}}
\def\EMF25{\ensuremath{\Ebar^{\perp}_{25}}}
\def\ECM60{\ensuremath{\Ebar^b_{60}}}

\def\TKM25{\ensuremath{N_{tk}^{25}}}

\def\snu{\ifmath{\tilde{\nu}}}

\def\slepr{\ifmath{\tilde{\ell}_R}}

\def\simge{\ \lower -2.5pt\hbox{$>$} \hskip-8pt \lower 2.5pt \hbox{$\sim$}\ }
\def\simle{\ \lower -2.5pt\hbox{$<$} \hskip-8pt \lower 2.5pt \hbox{$\sim$}\ }

\def\gappeq{\mathrel{\rlap {\raise.4ex\hbox{$>$}}
{\lower.6ex\hbox{$\sim$}}}}
\def\lappeq{\mathrel{\rlap{\raise.4ex\hbox{$<$}}
{\lower.6ex\hbox{$\sim$}}}}
\newcommand {\Be}{\begin{equation}}
\newcommand {\Ee}{\end{equation}}
\def\chna   {\mathrm{\tilde{\chi}_1^0}}
\def\qst    {\mathrm{\tilde{t}_1}}
\def\qsb    {\mathrm{\tilde{b}_1}}
\def\qstr   {\mathrm{\tilde{t}_R}}

\def\qast   {\mathrm{\bar{\tilde{t}}_1}}
\def\qasb   {\mathrm{\bar{\tilde{b}}_1}}
\def\snu    {\mathrm{\tilde{\nu}}}

\newcommand {\evis}{$E_{{vis}}$}

\def\snu{\ifmath{\tilde{\nu}}}

\def\susy#1{\ensuremath{\tilde{\mathrm{#1}}}}%
\def\slepton   #1{\ensuremath{\susy{\ell}^{#1}}}
\def\neutralino#1{\ensuremath{\susy{\chi}_{#1}^0}}
\newlength{\capindent}
\newlength{\capwidth}
\newlength{\figwidth}
\newcommand{\icaption}[2][!*!,!]{\hspace*{\capindent}%
  \begin{minipage}{\capwidth}
    \ifthenelse{\equal{#1}{!*!,!}}%
      {\caption{#2}}%
      {\caption[#1]{#2}}
  \end{minipage}}
\begin{document}
\begin{titlepage}
\title{Search for Scalar Leptons and Scalar Quarks at LEP}

\author{The L3 collaboration}

%
% The abstract
%
\begin{abstract}
Scalar partners of quarks and leptons, predicted in
supersymmetric models, are searched for in $\epem$ collisions at
centre-of-mass energies between 192 \gev{} and 209 \gev{} at LEP.
No evidence for any such particle is found in a data sample of
450 pb$^{-1}$. Upper limits on their production cross sections are set and
lower limits on their masses are derived in the framework of the Minimal
Supersymmetric Standard Model.
\end{abstract}

\submitted

\end{titlepage}

%
%
%%%%%%%%%%%%%%%%%%%%%%%%%%%%%%%%%%%%%%%%%%%%%%%%%%%%%%%%%%%%%%%%%%%%%%%%%%%%%%%
% Introduction
%%%%%%%%%%%%%%%%%%%%%%%%%%%%%%%%%%%%%%%%%%%%%%%%%%%%%%%%%%%%%%%%%%%%%%%%%%%%%%%
%
\section{Introduction}

The Minimal Supersymmetric extension of the Standard Model (MSSM)~\cite{thsusy,mssm2}
postulates a scalar partner,
$\mathrm{ \tilde {f}_{L,R} }$, for each weak eigenstate of Standard Model
(SM) fermions $\mathrm{f_{L,R}}$. Generally, the left, 
$\mathrm{\tilde{f}_{L} }$, and right,
$\mathrm{\tilde{f}_{R}}$, eigenstates mix to form mass
eigenstates. This mixing is an unitary transformation of the
$\mathrm{\tilde{f}_{R}}$ and $\mathrm{\tilde{f}_{L}}$ states,
parameterised by a mixing angle, $\theta_{\rm LR}$. Since the
off-diagonal elements of the sfermion mass matrix are proportional
to the SM partner mass, the mixing is expected to be relevant only for
scalar fermions of the third family: the scalar top, $\susy{t}_{\rm L,R}$, the scalar
bottom, $\susy{b}_{\rm L,R}$, and the scalar tau, $\susy{\tau}_{\rm L,R}$. The lightest scalar quarks are
denoted as  $\susy{t}_1$ and $\susy{b}_1$.

The R-parity is a quantum number which distinguishes SM particles
from supersymmetric particles. If R-parity is conserved, 
supersymmetric particles are
pair-produced and the lightest supersymmetric particle, assumed
hereafter to be the lightest neutralino, \neutralino{1}, is stable.
In addition, the \neutralino{1} is weakly-interacting and hence
escapes detection. R-parity conservation is assumed in the
following, which implies that the 
the decay chain of
pair-produced supersymmetric particles always contains, besides the
relevant SM particles, at least two invisible neutralinos.
The typical signature of the production of scalar leptons and
scalar quarks is the presence of leptons or jets in events with
missing energy and momentum. The difference between the masses of the
scalar fermion and the \neutralino{1}, $\Delta M$, determines the
kinematic of the event.

The pair-production of scalar fermions in $\rm e^+e^-$ interactions
proceeds through the
$s$-channel $\gamma$ or $\mathrm{Z}$ exchange. For scalar
electrons, the production cross section is typically enhanced by the
$t$-channel exchange of a neutralino.

At LEP energies, all scalar fermions, but the scalar top, decay into
their SM partners mainly via $\susy{f} \rightarrow \neutralino{1} \rm
f$. Cascade decays, such as $\susy{f} \rightarrow \neutralino{2} \rm f
\ra \neutralino{1}\Zzv \rm f $ are also possible and may dominate in
some regions of the MSSM parameter space. According to the values of
the scalar top mass and couplings, four channels can become dominant
among the possible scalar top decays: $\qst \to \mathrm{c} \chna$,
$\mathrm{b}\nu_{\ell}\tilde{\ell}$, $\mathrm{b}\ell\tilde{\nu_{\ell}}$
and $\mathrm{b}\tilde{\chi}^{+}_1$. The additional decay into
$\mathrm{b}\tilde{\chi}^{0}_1\rm f \bar{f}'$ which can originate six-fermion final states is not considered~\cite{Boehm}. This topology is indirectly covered by searches in the framework of R-parity violation, which revealed no excess~\cite{l3rpv}.
In the following, for the $\susy{t}
\rightarrow \susy{\nu} \rm b \rm \ell$ decay, scalar neutrinos are
assumed to be lighter than charged scalar leptons. For this decay,
$\Delta M$ refers to the mass difference between the scalar top and
scalar neutrino masses.

The supersymmetric partners of the
right-handed leptons, $\rm\slepton{}_R$, are generally expected to be lighter
than their left-handed counterparts and are considered in the
following.  
If the mass difference between the right-handed scalar electron
and the lightest neutralino is very small the search for
$\rm\epem \rightarrow \susy{e}_{R}\susy{e}_{R}$ has little
sensitivity. The $\rm\epem
\rightarrow \susy{e}_{R}\susy{e}_{L}$ process is then considered. 
The left-handed
scalar electron, too heavy to be produced in pairs, decays into an
energetic electron, while the electron from the right-handed scalar
electron decay remains often invisible, leading to a
`single electron' topology.

Scalar leptons and scalar quarks are searched for at
centre-of-mass energies, $\sqrt{s}$, up to 209 GeV. The present
study supersedes previous L3 limits on scalar lepton~\cite{pubsle}
and scalar quark production~\cite{pubsq} obtained at lower
$\sqrt{s}$. Searches for scalar fermions were also
reported by other experiments at LEP~\cite{limits} and at the 
TEVATRON~\cite{CDFD0}.
Table~\ref{topologies} summarises the investigated processes and decay modes 
together with the studied topology. 

\begin{table}[h!]
\begin{center}
\begin{tabular}{|rcl|rcl|c|}
  \hline
\multicolumn{3}{|c}{Process} & \multicolumn{3}{|c|}{Decay mode} & Topology \\
  \hline
  \hline
\rule{0pt}{14pt}$ \epem  $& $\rm\rightarrow $& $\susy{\ell}_{R} \bar{\susy{\ell}}_{R}$ & $\rm\susy{\ell}_{R} $& $\rightarrow $& $\neutralino{1} \rm \ell$ & Acoplanar leptons  \\
\rule{0pt}{14pt}$ \epem  $& $\rm\rightarrow $& $\susy{e}_{R} \susy{e}_{L}$             & $\rm\susy{e}_{L,R}  $& $\rightarrow $& $\neutralino{1} \rm e $ & Single electron \\ 
\rule{0pt}{14pt}$ \epem  $& $\rightarrow $& $\susy{b} \bar{\susy{b}}$               & $\susy{b}        $& $\rightarrow $& $\neutralino{1} \rm b$& Acoplanar b-jets \\
\rule{0pt}{14pt}$ \epem  $& $\rightarrow $& $\susy{t} \bar{\susy{t}}$               & $\susy{t}        $& $\rightarrow $& $\neutralino{1} \rm c$& Acoplanar jets \\
\rule{0pt}{14pt}$ \epem  $& $\rightarrow $& $\susy{t} \bar{\susy{t}}$               & $\susy{t}        $& $\rightarrow $& $\susy{\nu} \rm b \ell$& Acoplanar jets and leptons \\
\rule{0pt}{14pt}$ \epem  $& $\rightarrow $& $ \susy{q} \bar{\susy{q}}$              & $\susy{q}        $& $\rightarrow $& $\neutralino{1} \rm{q} $& Acoplanar jets \\
  \hline
\end{tabular}
\caption{Summary of the investigated processes, decay modes and studied topologies.
\label{topologies}}
\end{center}
\end{table}

%%%%%%%%%%%%%%%%%%%%%%%%%%%%%%%%%%%%%%%%%%%%%%%%%%%%%%%%%%%%%%%%%%%%%%%%%%%%%%
%
% Signal and Background
%
%%%%%%%%%%%%%%%%%%%%%%%%%%%%%%%%%%%%%%%%%%%%%%%%%%%%%%%%%%%%%%%%%%%%%%%%%%%%%%%
%
\section{Data samples and Monte Carlo simulation }
The data used in the present analysis were collected with the L3
detector~\cite{L3} at LEP and correspond to an integrated
luminosity of $450.5 \rm pb^{-1}$ at $\sqrt{s}=192-209\GeV$. Two
average centre-of-mass energies are considered in the following:
$196\GeV$ and $205\GeV$, with corresponding integrated luminosities
of $233.2 \rm pb^{-1}$ and $217.3 \rm pb^{-1}$.

SM processes are simulated with the following Monte Carlo (MC)
generators: {\tt PYTHIA}~\cite{pythia} for
  $\ee \rightarrow \mathrm{q\bar{q}(\gamma)}$,
  $\ee \rightarrow \mathrm{Z}\,\ee$ and
  $\ee \rightarrow \mathrm{Z}\mathrm{Z} $,
{\tt EXCALIBUR}~\cite{excali} for
   $\ee \rightarrow \mathrm{W^\pm\, e^\mp \nu}$,
{\tt KORALZ}~\cite{koralz} for
   $\ee \rightarrow \mu^+\mu^-(\gamma)$ and
   $\ee \rightarrow \tau^+\tau^-(\gamma)$,
{\tt BHWIDE}~\cite{bhwide} for
   $\ee \rightarrow \ee(\gamma)$ and
{\tt KORALW}~\cite{koralw} for
   $\ee \rightarrow \mathrm{W^+ W^-}$.
Two-photon interaction processes are simulated using {\tt
DIAG36}~\cite{diag36} for $\ee \rightarrow \ee \ell^+\ell^-$ and
{\tt PHOJET}~\cite{phojet} for $\ee \rightarrow \ee\,
\mathrm{hadrons}$, requiring at least 3 \gev{} for the invariant
mass of the two-photon system. The number of simulated events for
each background process is  more than 100 times the
data statistics, except for two-photon processes for which the
MC statistics amounts to about 7 times that of the data.

Signal events for scalar leptons are generated with the 
{\tt SUSYGEN}~\cite{susygen} MC program, for scalar lepton masses, $M_{\susy {\ell}}$, ranging from 45 \gev{} up to the kinematic
limit, 
and for values of $\dm$ varying between 3
\gev{} and $M_{\susy {\ell}}-1 \GeV$. For scalar quarks, a generator~\cite{sqgen} based
on {\tt PYTHIA} is used. Scalar quark masses vary from
45 \gev{} up to the kinematical limit and $M_{\neutralino{1}}$ varies from 1
\gev{} to $M_{\qst} - 3 \gev$ and to $M_{\qsb} - 7 \gev$, for
scalar top and bottom respectively. The $\qst \to \mathrm{b} \ell
\snu$ and $\qst \to \mathrm{b} \tau \snu$ channels are generated
with $\snu$ mass ranging from the $43\gev$ limit~\cite{LEP1snu} up to $M_{\qst} - 8 \gev$.
%48+27+27+47
In total, about 180 samples are generated, each with at least 1000
events.
 
The response of the L3 detector is simulated using the
{\tt GEANT} package~\cite{geant}. It takes into account effects of
energy loss, multiple scattering and showering in the detector
materials and in the beam pipe. Hadronic interactions  are
simulated with the {\tt GHEISHA} program ~\cite{geisha}. Time-dependent detector inefficiencies are monitored during data taking
and reproduced in the simulation.

%%%%%%%%%%%%%%%%%%%%%%%%%%%%%%%%%%%%%%%%%%%%%%%%%%%%%%%%%%%%%%%%%%%%%%%%%%%%%
%
% Analysis procedure
%
%%%%%%%%%%%%%%%%%%%%%%%%%%%%%%%%%%%%%%%%%%%%%%%%%%%%%%%%%%%%%%%%%%%%%%%%%%%%%
\section{Event selection}

\subsection{Analysis procedure }

Besides the common signature of missing momentum in the direction
transverse to the beam axis, signals from supersymmetric particles are
further specified according to the number of leptons or the
multiplicity of hadronic jets in the final state.

Signatures of scalar leptons are simple and the final states mostly
contain just two acoplanar leptons of the same generation. To account
for the three lepton flavours, three different selections are
performed. For scalar electrons and muons a pair of electrons or muons
is required in the event, respectively,  while scalar taus are selected as low
multiplicity events with electrons or muons or with narrow
jets. Events from the $\qst \to \mathrm{c} \chna$ and $\qsb \to
\mathrm{b} \chna$ processes contain two high-multiplicity acoplanar
jets originated by c or b quarks. In addition, two charged leptons are
present when both scalar top quarks decay via $\qst \to \mathrm{b} \ell
\snu$.

An optimization procedure is devised~\cite{pubsle} which maximizes
signal efficiency and background rejection by varying simultaneously
all cuts for a given process.  The signal topology depends on
$\Delta{M}$ and therefore the optimization is repeated for different
values of $\Delta{M}$.  Details of the selections performed for each
topology are given in the following.

\subsection{Acoplanar leptons}

Scalar leptons are searched for in events with
two isolated leptons of the same flavour. The lepton 
identification and isolation criteria follow those used at
lower $\sqrt{s}$ \cite{paper183}. An electron is isolated if the
calorimetric energy deposition in a $10^{\circ}$ cone around its
direction is less than 2 \gev{}. Muon isolation requires an
energy below 2 \gev{} in the cone between $5^{\circ}$ to
$10^{\circ}$ around the muon direction. A tau is isolated if the
energy deposition in the cone between $10^{\circ}$ and
$20^{\circ}$ around its direction is less than 2 \gev{} and less
than 50\% of the tau energy. Furthermore, the energy deposition in
a cone between $20^{\circ}$ and $30^{\circ}$ must be less than 60\%
of the tau energy.

The large background from two-photon interactions is rejected with
cuts on the lepton transverse momentum, the visible mass, $M_{vis}$,
the transverse missing momentum, $P^{miss}_T$, the energy deposited at
low polar angle, $E_{30}$, and the sine of the polar angle of the
missing momentum, $\sin\theta_{miss}$.  Acoplanarity and acollinearity
cuts together with upper bounds on the visible energy, $E_{vis}$,
reduce the background from W boson and fermion pair-production. After
these cuts, the distributions of selection variables for data and
Monte Carlo are in good agreement, as shown in
Figure~\ref{fig:presel}a for the energy of the most energetic lepton, 
$E_{1}$.

The final selections are optimised for each scalar lepton flavour,
using a set of parameterized cuts ($E_{vis}$, $P^{miss}_T$,
$M_{vis}$, $E_{1}$) together with fixed cuts (acoplanarity,
acollinearity and $\sin \theta_{miss}$). The parameterised cuts
depend on $Z = (\Delta{M}/M_{\tilde{\ell}})\times{E_{beam}}$, to
reflect the dependence of the final state topologies on
$\Delta{M}$ and $M_{\tilde{\ell}}$. $E_{beam}$ is the beam energy.
The variables used for each
selection are described in Reference~\citen{pubsle}.

The selection efficiencies for scalar lepton pair-production, the number
of candidates in data and the SM expectations are given in
Table~\ref{slepton} for three $\Delta{M}$ regions.

\subsection{Single electron}

The single-electron analysis requires one or two identified electrons.
Cuts on $E_{vis}$ and $\sin\theta_{miss}$ are applied in order to
reject background from two-photon interactions. At least one electron
with energy greater than $5\GeV$ is required. The electron energy has
to be less than 65 \GeV\ to reject photon conversion from the $\rm
e^+e^- \rightarrow \nu\bar{\nu}\gamma$ process when the two tracks are
not resolved. If two electrons are selected, their acoplanarity must be
between $10^{\circ}$ and $160^{\circ}$ and the energy of the second
electron must be less than $5\GeV$ to suppress background from W
pair-production. To remove events with additional activity in the
detector, the difference between the total energy and the energy of
the most energetic electron
must be less than $5\GeV$. In addition, a cut $P^{miss}_T > 15
\GeV$ is applied. If no second electron of at least $100\MeV$ is
detected, this cut is released to $P^{miss}_T > 10 \GeV$.
Figure~\ref{fig:presel}b compares data and MC for the energy of the
most energetic electron, the remaining background originates from
four-fermion final states.  Signal efficiencies vary from 3\% at
$\Delta M=M_{\rm\susy{e}_L} - M_{\neutralino{1}} = 5 \GeV $ up to 60\%
for $\Delta M=60 \GeV$.

\subsection{Acoplanar jets}
\begin{figure}
  \begin{center}
    \begin{tabular}{cc}
      \includegraphics*[width=0.4\textwidth]{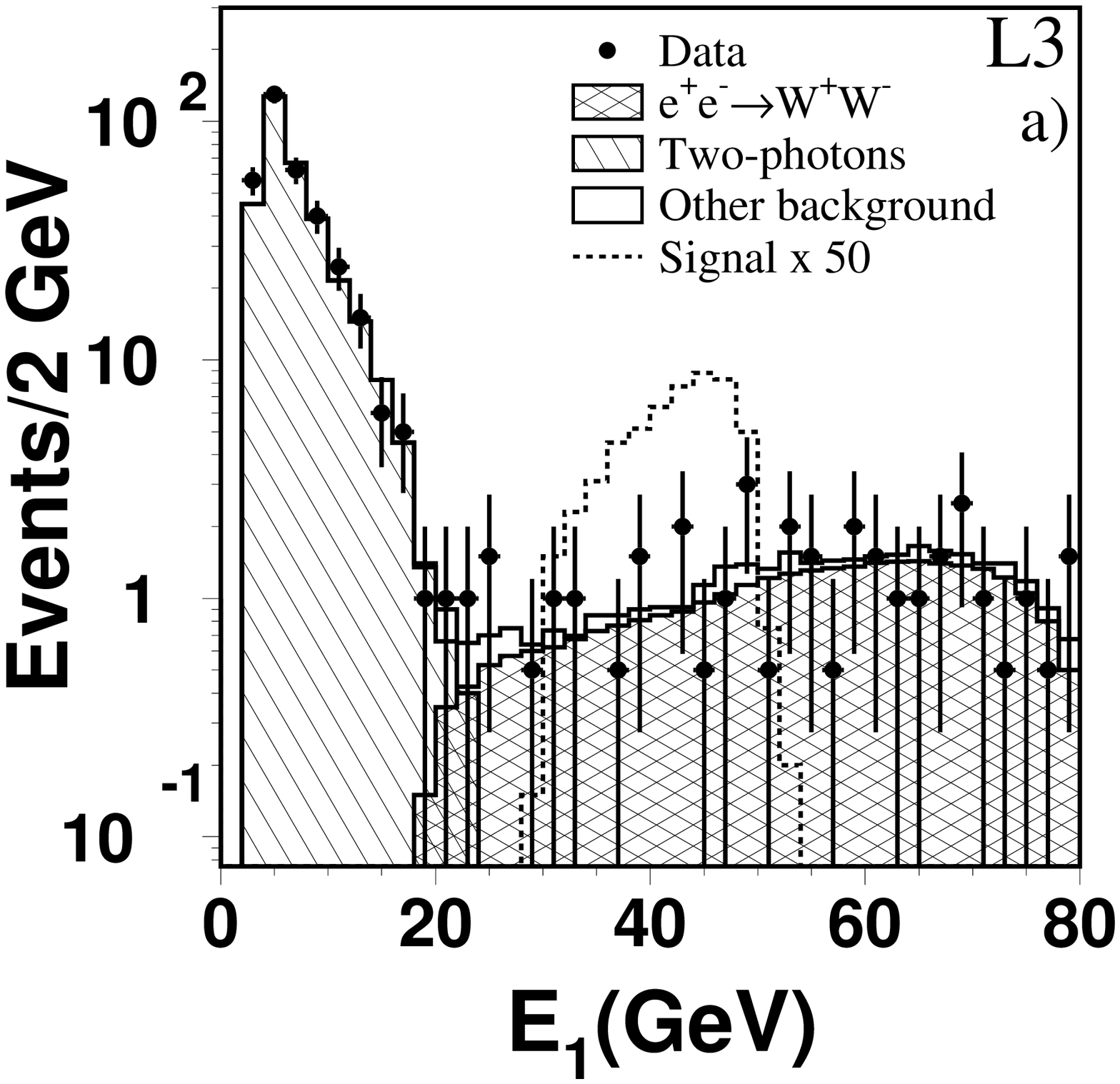}&
      \includegraphics*[width=0.4\textwidth]{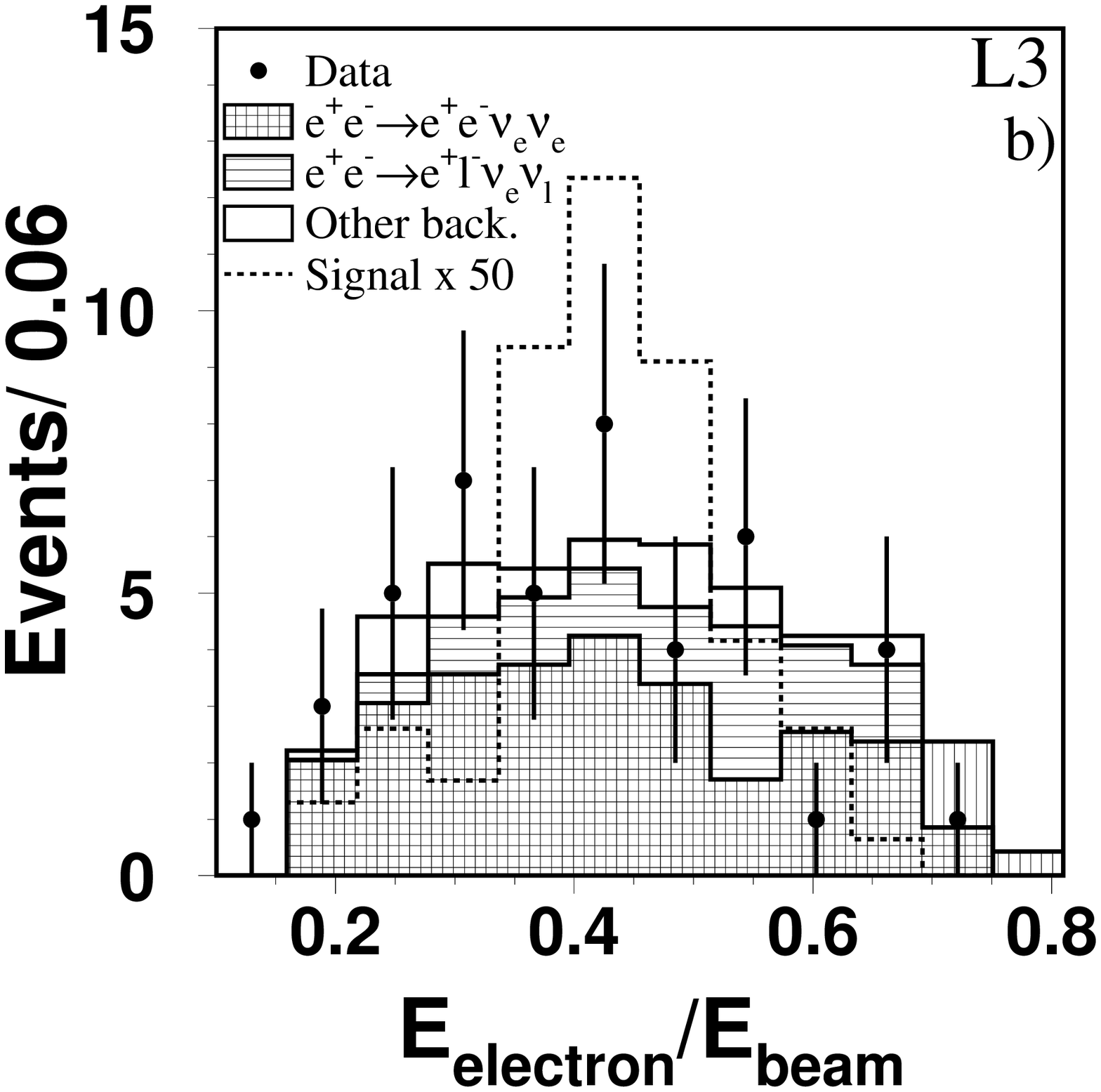}\\
      \includegraphics*[width=0.4\textwidth]{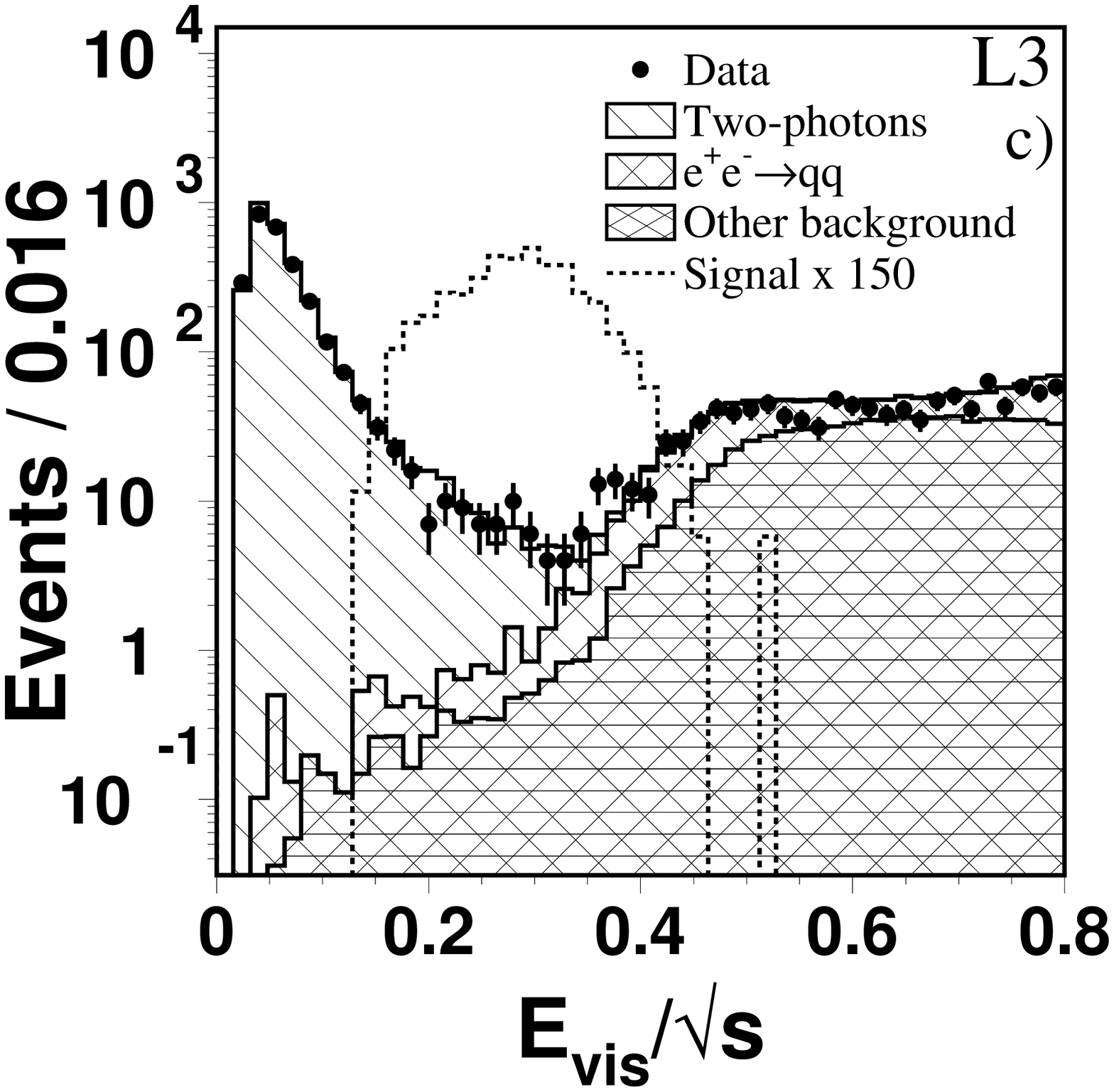}&
      \includegraphics*[width=0.4\textwidth]{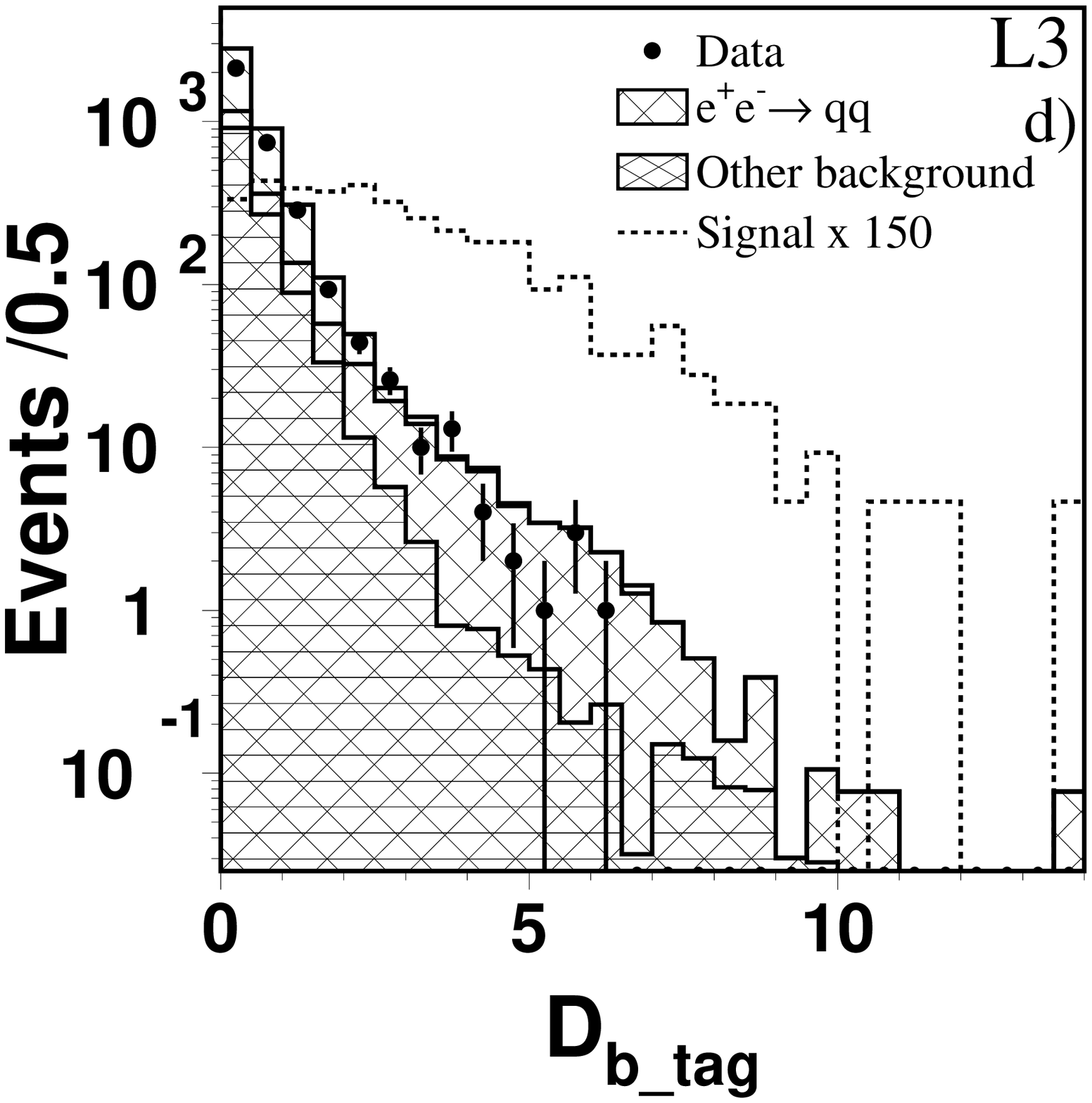}\\
    \end{tabular}
  \end{center}
\caption { \label{fig:presel} Distributions in data and MC of the energy of the most energetic lepton
  of the a) scalar lepton searches and b) single electron analysis. c)
  visible energy and d) b-tag variable for the scalar quark
  analysis. Signal events are scaled by the factors indicated in the figures and
  correspond to a) $M_{\susy{\ell_{\rm R}}} = 90\gev$ and
  $M_{\neutralino{1}} = 40\gev$, b)  $M_{\susy{\rm e}_{\rm L}} = 110\GeV$ and
  $M_{\neutralino{1}} = 50\gev$, c) and d) $\qst
\to \mathrm{c} \chna$ decay for 
$M_{\qstr}=90 \gev$, $M_{\chna}=60 \gev$.
}

\end{figure}

The search for scalar quarks decaying  into quarks and neutralino is
based on events with two high-multiplicity acoplanar jets.
The DURHAM algorithm~\cite{durham} is used for the clustering of hadronic jets.
A common preselection is applied~\cite{pubsq} which  is based on:
\evis, the calorimetric cluster multiplicity, $P{{^{miss}_T}}$,
$E_{30}$ and $\sin\theta_{miss}$. After this preselection, the data
agree well with the SM expectations, as depicted in
Figures~\ref{fig:presel}c and~\ref{fig:presel}d.

Four selections are optimised for scalar top quarks and four for scalar
bottom quarks. They depend on  $\Delta M$ and cover
the regions  $5-10\gev$, $10-20\gev$, $20-40
\gev$ and above $40\gev$. Lower cuts on \evis/$\sqrt{s}$
and $P{^{miss}_T}/\sqrt{s}$~ separate the signal from the two-photon
background, whereas an upper cut on \evis/$\sqrt{s}$  removes 
events from four-fermion final states.
A cut on $\sin\theta_{{miss}}$ also rejects the two-photon
background.
Cuts on the jet widths and on the absolute value of the projection of
the total momentum of the jets onto the direction perpendicular to
thrust, computed in the transverse plane, further suppress the
two-photon as well as $\mathrm{W^+W^-}$ and
$\mathrm{q\bar{q}(\gamma)}$ backgrounds.

For the scalar bottom selection, b-quark identification in the
final state is enforced by an additional cut on the event b-tagging
variable~\cite{pubsq}, $D_{b-tag}$.

The expected signal efficiencies at various $\Delta M$ values are given in
Table~\ref{squark} together with the observed number of events and the
SM background expectations.

\subsection{Acoplanar jets and leptons }

A selection of events with two acoplanar jets and one or two isolated leptons
complements the scalar top searches in presence of the 
$\qst \to \mathrm{b} \ell \snu$ decay. Large values of the $D_{b-tag}$
variables are required  for the two jets and additional cuts on
\evis/$\sqrt{s}$ reject part of the two-photon and four-fermion events.
Lower cuts on the energy of the leptons 
suppress background from two-photon interactions at low $\Delta M$ and
the $\mathrm{q\bar{q}(\gamma)}$ final state at medium $\Delta M$. At
high $\Delta M$, an upper cut on the lepton energy reject
four-fermion events. This selection covers the $\Delta M$ region above the limit $M_{\snu}
>43 \gev$.

The expected signal efficiencies for scalar top detection are
given in Table~\ref{squark} together with data counts and the SM
background expectations, for various $\Delta M$ values.

\section{Results}

\subsection{Cross section limits}

As discussed above and summarized in Table~\ref{results}, no
excess with respect to the Standard Model expectations is observed in the data. Upper limits on the production cross
section are therefore derived combining these results with those
obtained at lower $\sqrt{s}$~\cite{pubsle,pubsq}. This combination scales 
the signal cross sections
with $\sqrt{s}$ and the limits refer to $\sqrt{s}=205\GeV$.

\begin{figure}
  \begin{center}
    \begin{tabular}{cc}
      \includegraphics*[width=0.4\textwidth]{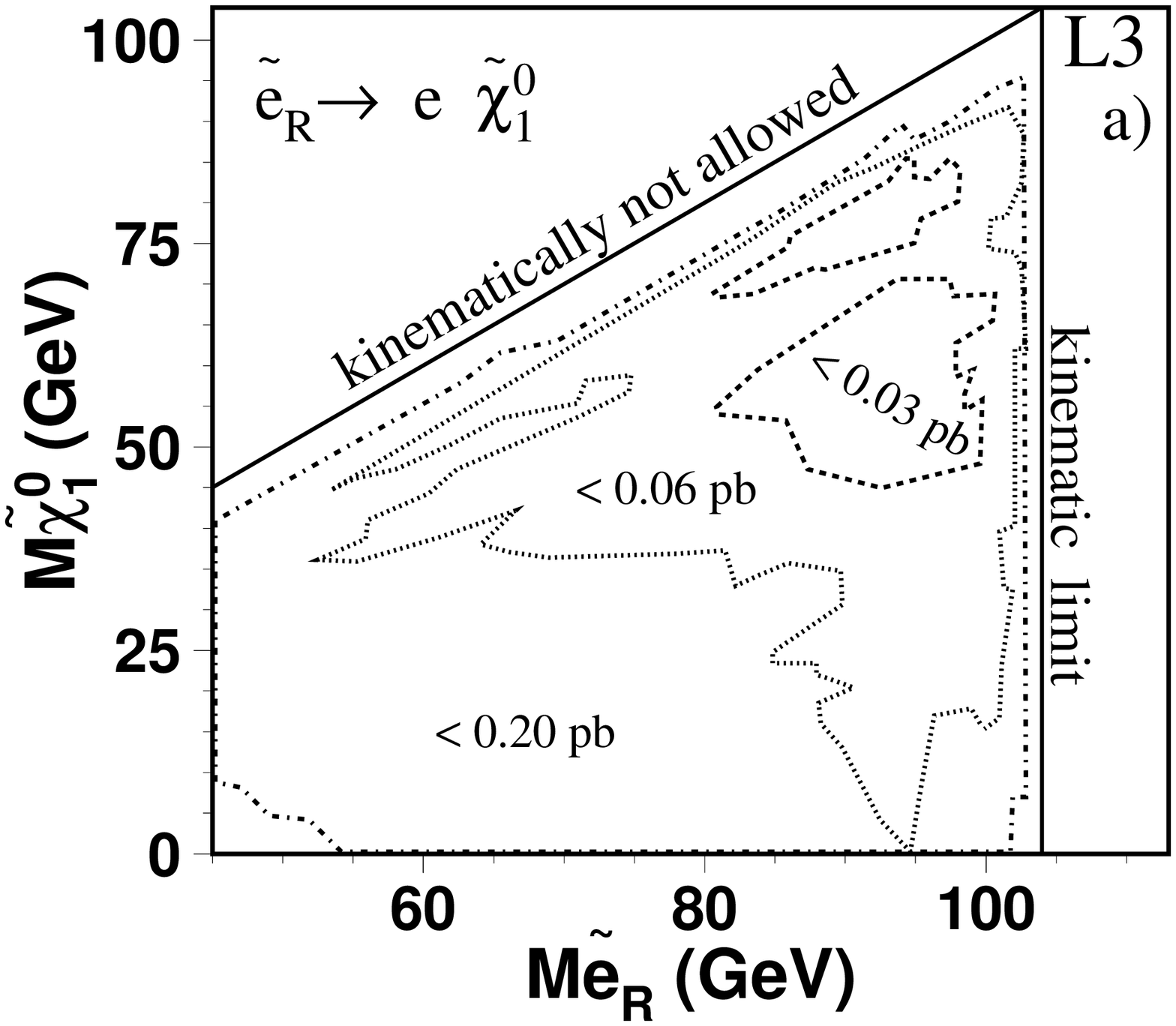}&
      \includegraphics*[width=0.4\textwidth]{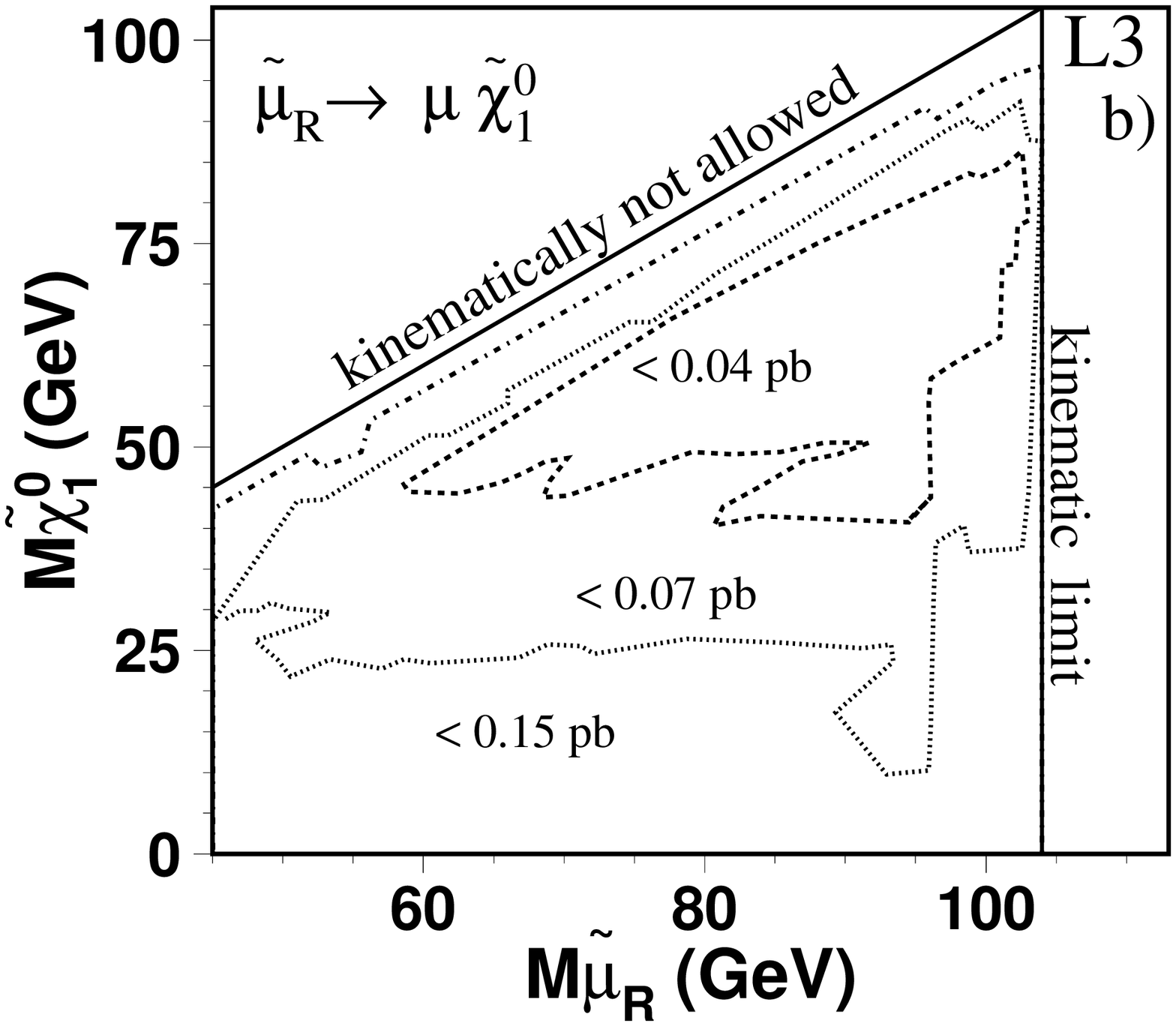}\\
      \includegraphics*[width=0.4\textwidth]{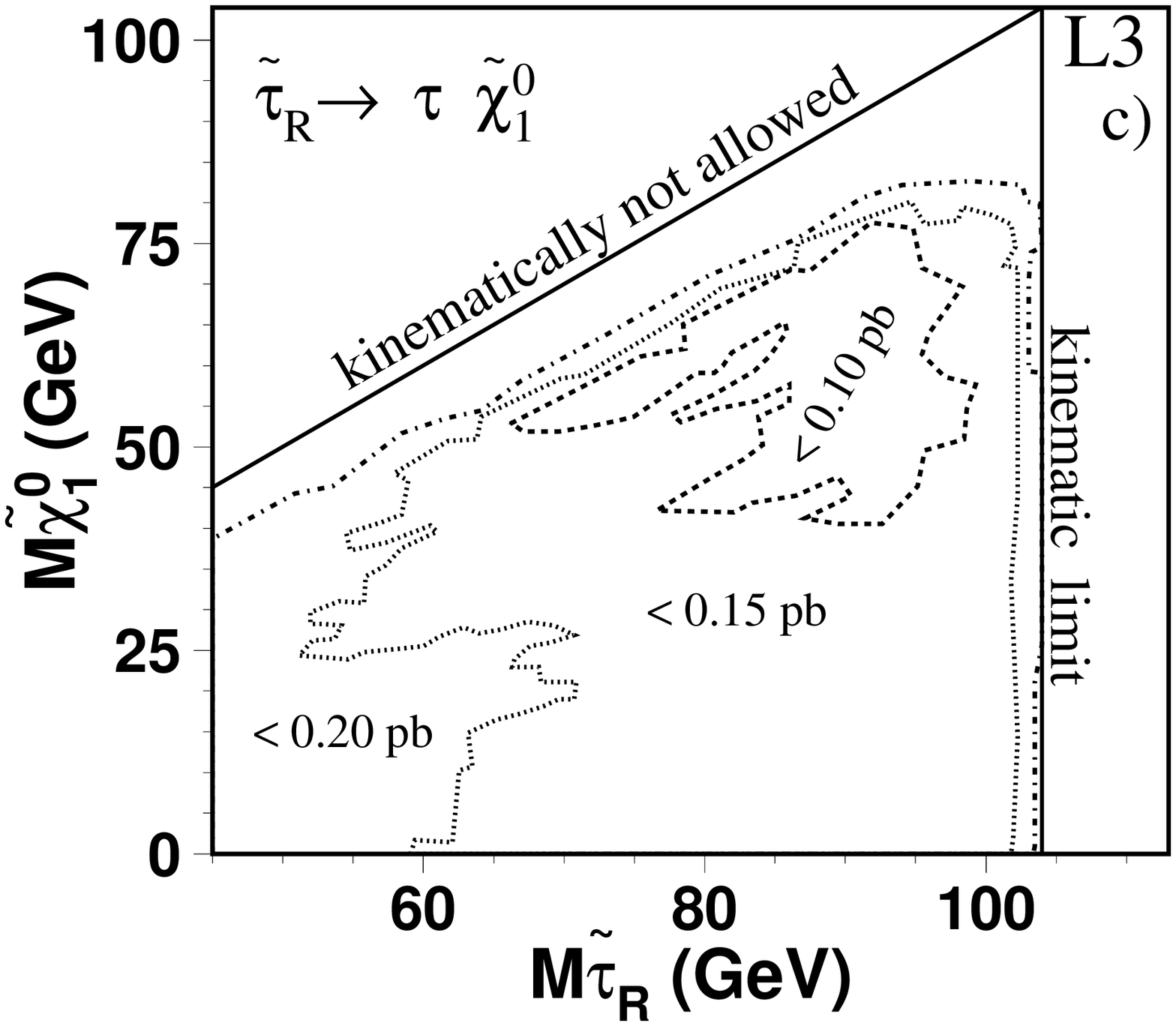}&\\
    \end{tabular}
  \end{center}
      \caption{ \label{slep-upp} Model independent upper limits on the
	$\epem  \rm\rightarrow \susy{\ell}_{R} \bar{\tilde{\ell}}_{R}$ 
	cross section
      in the $M_{\tilde{\chi}^0_1} - M_{\tilde{\ell}}$ plane, for a)
	scalar electrons, b) scalar muons and c) scalar taus.}

\end{figure}

Figures~\ref{slep-upp} and~\ref{fig:xbrexcl1} show the 95\% confidence
level (CL) upper bounds on the production cross sections as a function
of the scalar fermion masses and of the neutralino mass. The case of
right handed scalar leptons and of the lightest scalar quarks is
considered. These limits include~\cite{cousins} the systematics
effects discussed below.

\subsection{Systematic uncertainties}

Systematic uncertainties on the signal efficiency for scalar
lepton searches and on all background predictions are dominated by
Monte Carlo
statistics. They are smaller than 5\%. The main systematic
uncertainties on the scalar quark signal selection efficiency
arise from uncertainties on the production mechanism,
hadronisation and decay of the scalar quark~\cite{pubsq}.
These uncertainties are in the range from 7\% to 18\% for scalar top,
with the highest
uncertainty in the very low $\dm$ region. For scalar bottom, the highest uncertainty
is about 10\% and is observed in the very low and high $ \dm$ regions.

\section{Interpretations in the MSSM}

In the MSSM, with Grand Unification assumptions~\cite{mssm1}, the
masses and couplings of the supersymmetric particles as well as their
production cross sections are described~\cite{mssm2} in terms of 
five parameters: $\tan\beta$, the ratio of the
vacuum expectation values of the two Higgs doublets, $M_2 \simeq
0.81 \times m_{1/2}$, the gaugino mass parameter, $\mu$, the
higgsino mixing parameter, $m_0$,  the common mass for scalar
fermions at the GUT scale and $A_0$, the trilinear coupling in
the scalar fermion sector. We investigate the following MSSM parameter
space:
$$
   \begin{array}{rclcrcl}
         1 \leq& \tan\beta & \leq 60 ,     &&
0 \gev \leq &M_2 &\leq 2000 \gev ,\\
 -2000 \gev \leq& \mu       & \leq 2000 \gev,&&
0 \gev \leq& m_0         &\leq  500 \gev,
  \end{array}
$$
\vspace{-1.5em}
$$
 -1000\GeV<A_0<1000\GeV
$$
The limits on the production cross section for scalar leptons and
scalar quarks discussed above are translated into exclusion
regions in the MSSM parameter space. To derive these limits,
we optimise the event selection for each point in the MSSM parameter
space by choosing the combination of selections
which provides the highest sensitivity for each 
process. This sensitivity is derived by calculating at each point
the production cross sections and the decay branching fractions of
scalar leptons and scalar quarks. For the latter, the mixing angle
$\theta_{\rm LR}$ is also considered. A
point of the MSSM parameter space is excluded if any of these
calculated cross sections
exceeds its corresponding experimental limit. Mass lower limits are
derived as the lowest value for the mass of a
particle over all points which are 
not excluded.

\subsection{Limits on scalar lepton masses}

\begin{figure}
  \begin{center}
    \begin{tabular}{cc}
      \includegraphics*[width=0.4\textwidth]{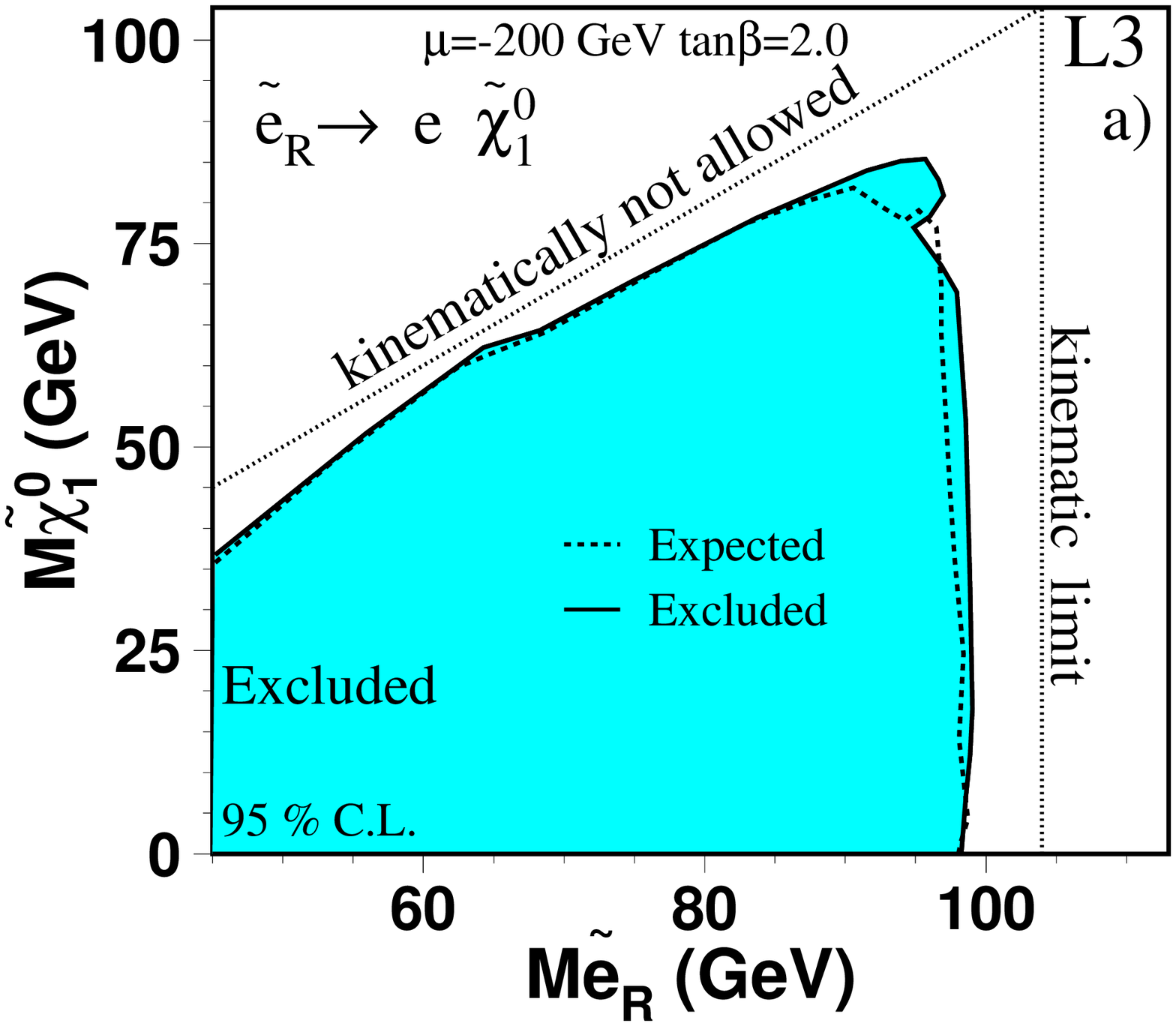}&
      \includegraphics*[width=0.4\textwidth]{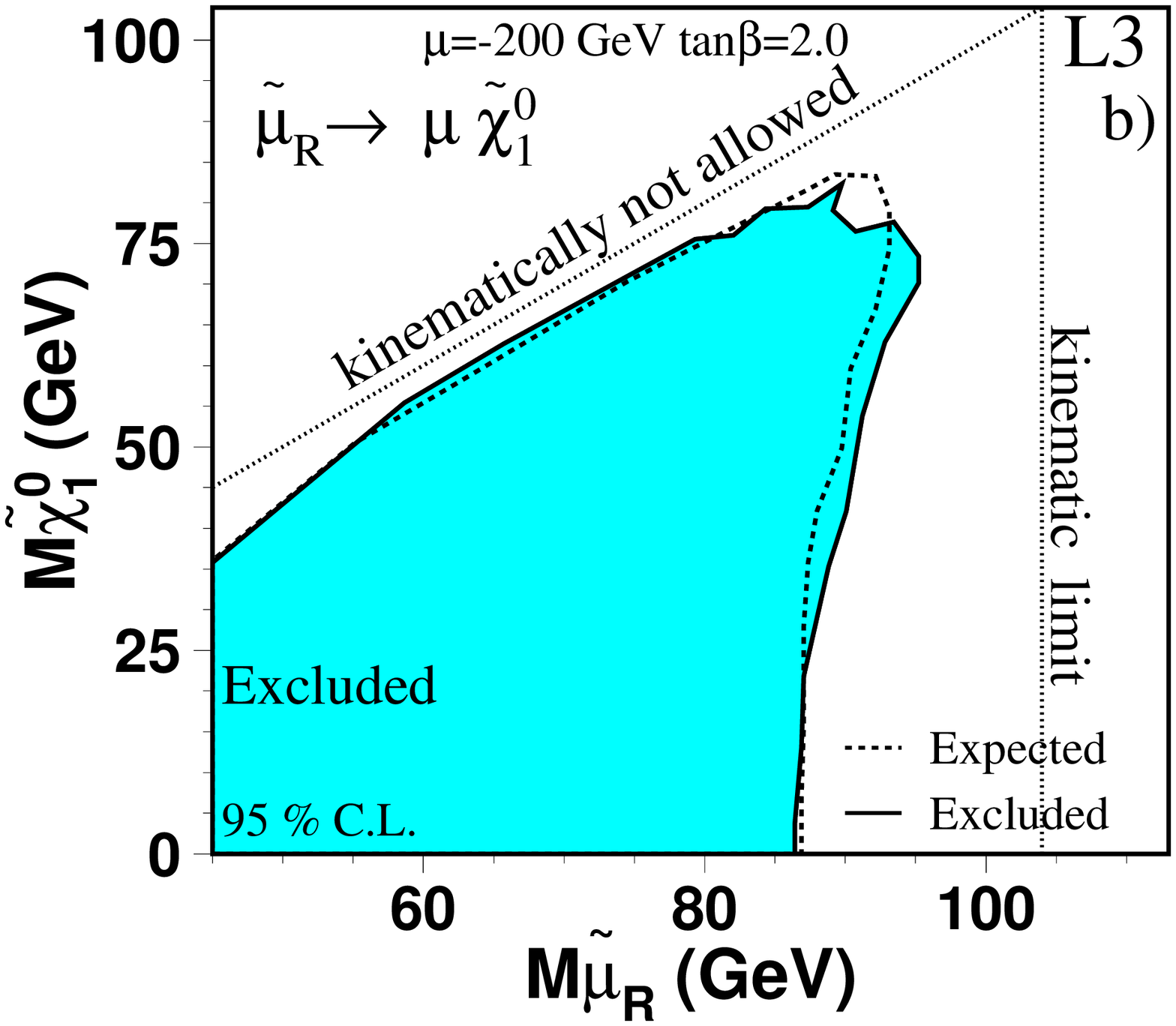}\\
      \includegraphics*[width=0.4\textwidth]{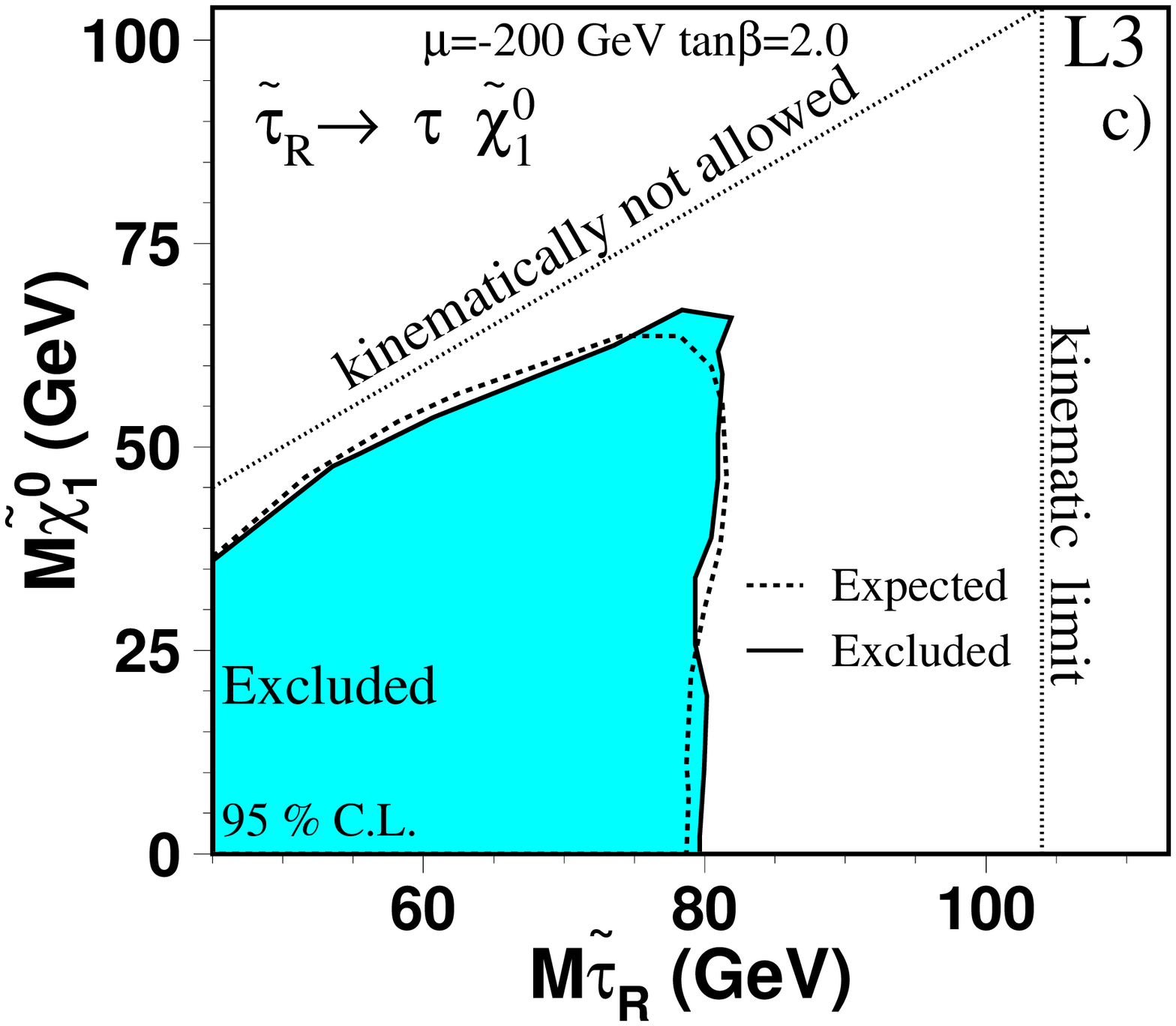}&\\
    \end{tabular}
  \end{center}
     \caption{\label{fig:rleptons} Regions of the plane
     $M_{\neutralino{1}} - M_{\susy{\ell}_{\rm R}}$ excluded in the
     MSSM for a)
     scalar electrons, b) scalar muons and c) scalar taus.}
\end{figure}

Figures~\ref{fig:rleptons}a, \ref{fig:rleptons}b and
\ref{fig:rleptons}c show the exclusion contours in the
$M_{\neutralino{1}} - M_{\rm\slepton{}_{\rm R}}$ plane obtained by
considering only the reaction $\rm\ee \rightarrow \tilde{\ell}_R
\bar{\tilde{\ell}}_R$ for $\mu = -200 \gev$ and $\tan\beta = 2$. These
exclusions hold for $\tan\beta \ge 2 $ and $|\mu| \ge 200$.

Under these assumptions, 95\% CL lower limits on the masses of 
scalar leptons are
derived as  $94.4 \gev$ for scalar electrons with $\dm
> 10 \gev$,  $86.7 \gev$ for scalar muons with $\dm
> 10 \gev$ and
 $78.3 \gev$ for scalar taus with $\dm >15 \gev$.

\begin{figure}
  \begin{center}
    \begin{tabular}{c}
      \includegraphics*[width=0.6\textwidth]{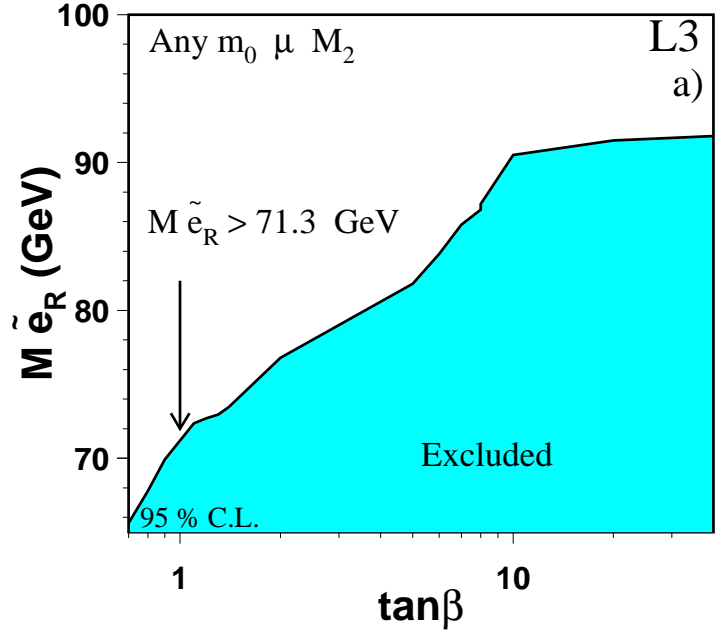}\\
      \includegraphics*[width=0.6\textwidth]{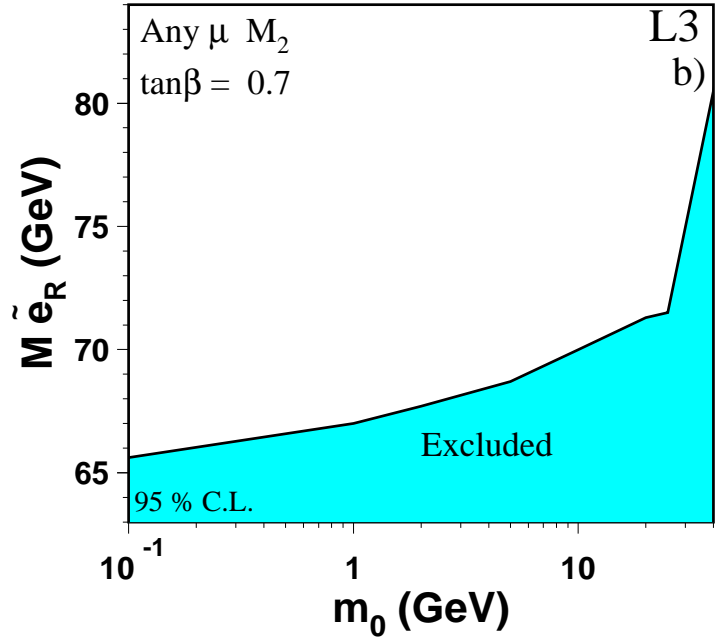}\\
    \end{tabular}
  \end{center}
\caption{\label{fig:single_electron} Absolute
$\mathrm{\tilde{e}_R}$ mass limit as a function of a) tan$\beta$ and b) $m_0$. }
\end{figure}

The limiting factor towards an absolute limit on the scalar
electron mass is the lack of detection efficiency for very small
\dm{} values. This is overcome, in the constrained MSSM, by 
using the $\rm\epem\ra\susy{e}_R\susy{e}_L$ process.
The searches for acoplanar electrons  and single electrons are
combined to derive a
 lower limit on $M_{\susy{e}_R}$ as a function of $\tan\beta$
and for any value of $m_0$, $M_2$ and $\mu$ as shown in
Figure~\ref{fig:single_electron}a. For $\tan\beta < 1 $ the mass
difference between $\susy{e}_L$ and $\susy{e}_R$ decreases, reducing
the sensitivity of the single electron search.
As an example,
Figure~\ref{fig:single_electron}b shows the limit as a function of
$m_0$ for a fixed value of  $\tan\beta$.
For $\tan\beta \geq 1 $, the 95\% CL lower 
limit for the lightest scalar electron, independent of the MSSM
parameters, is:
$$  M_{\rm\susy{e}_R} \geq  71.3 \gev. $$
Assuming a common mass for the scalar leptons at the GUT scale,
this limit holds for the lightest scalar muon,
$\susy{\mu}_R$, as well.

\subsection{Limits on scalar quark masses}

\begin{figure}
  \begin{center}
    \begin{tabular}{cc}
      \includegraphics*[width=0.4\textwidth]{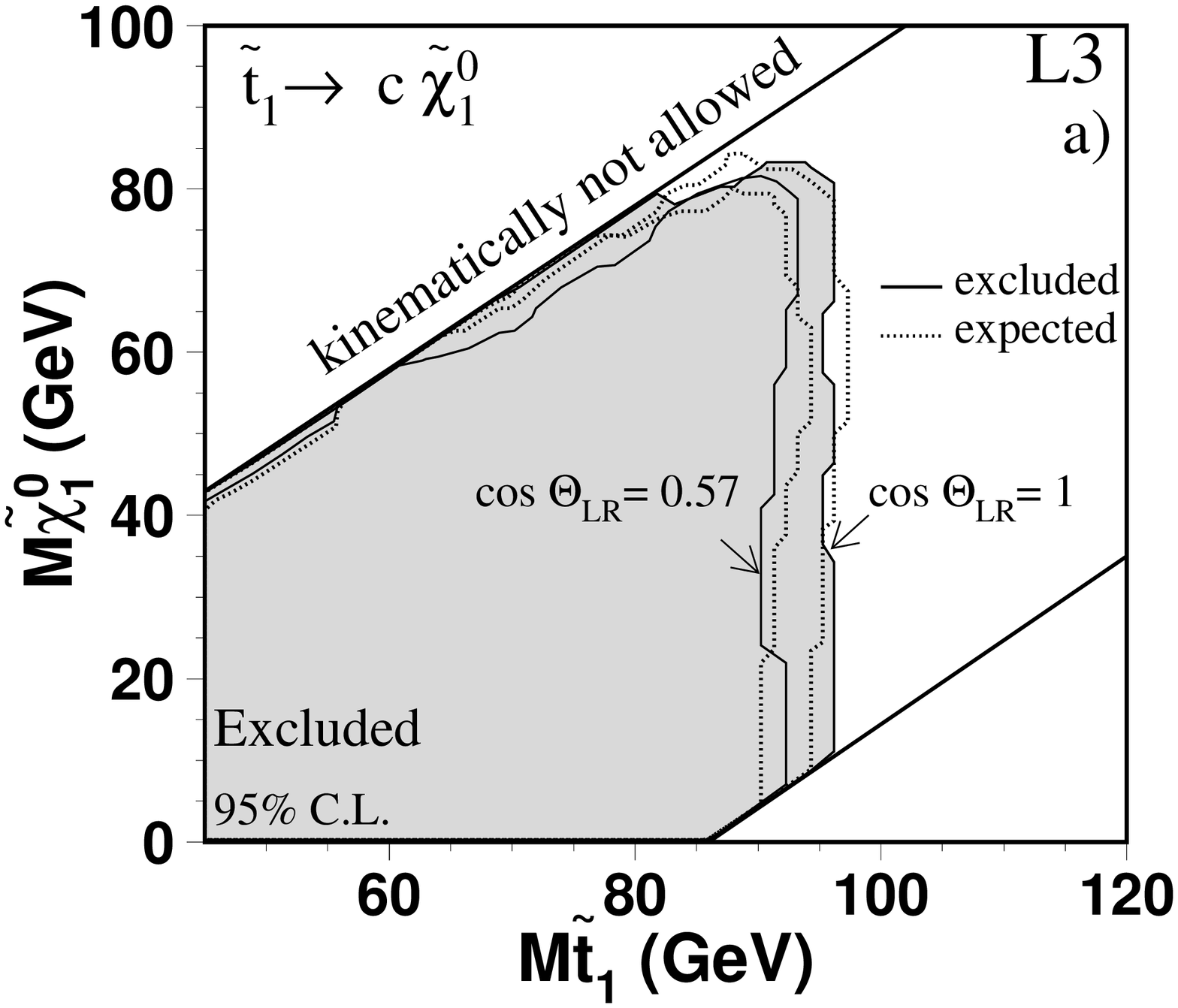}&
      \includegraphics*[width=0.4\textwidth]{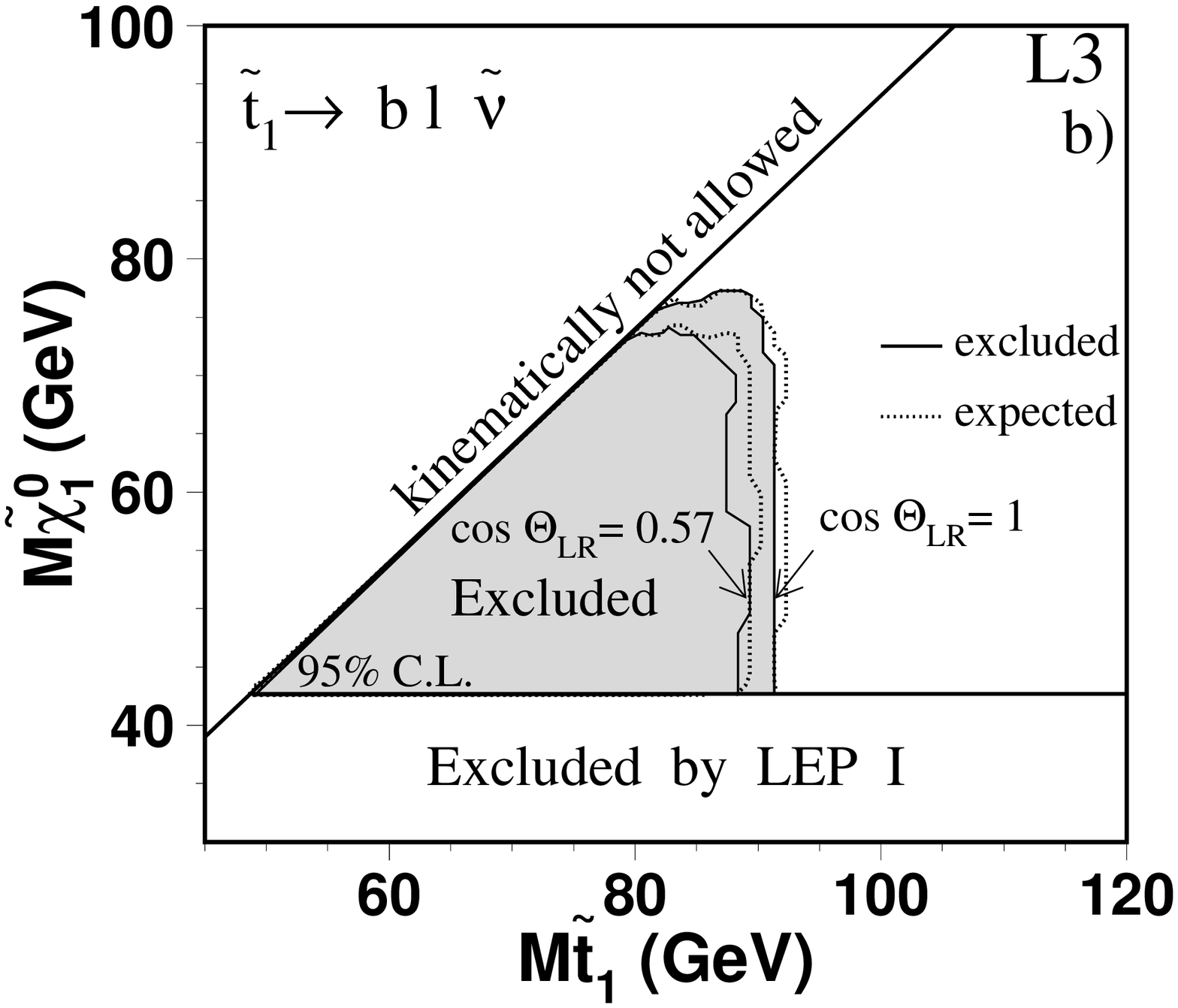}\\
      \includegraphics*[width=0.4\textwidth]{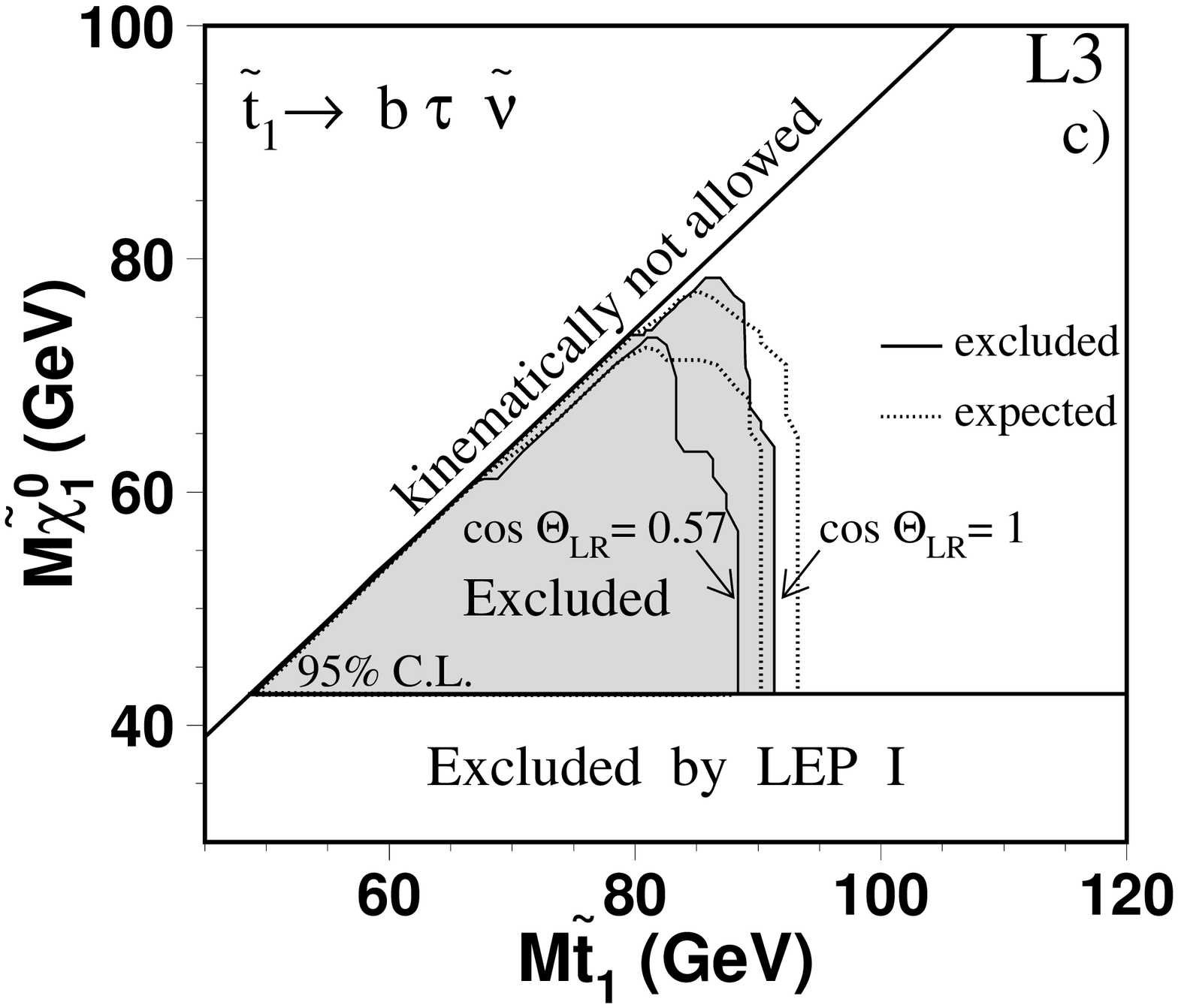}&
      \includegraphics*[width=0.4\textwidth]{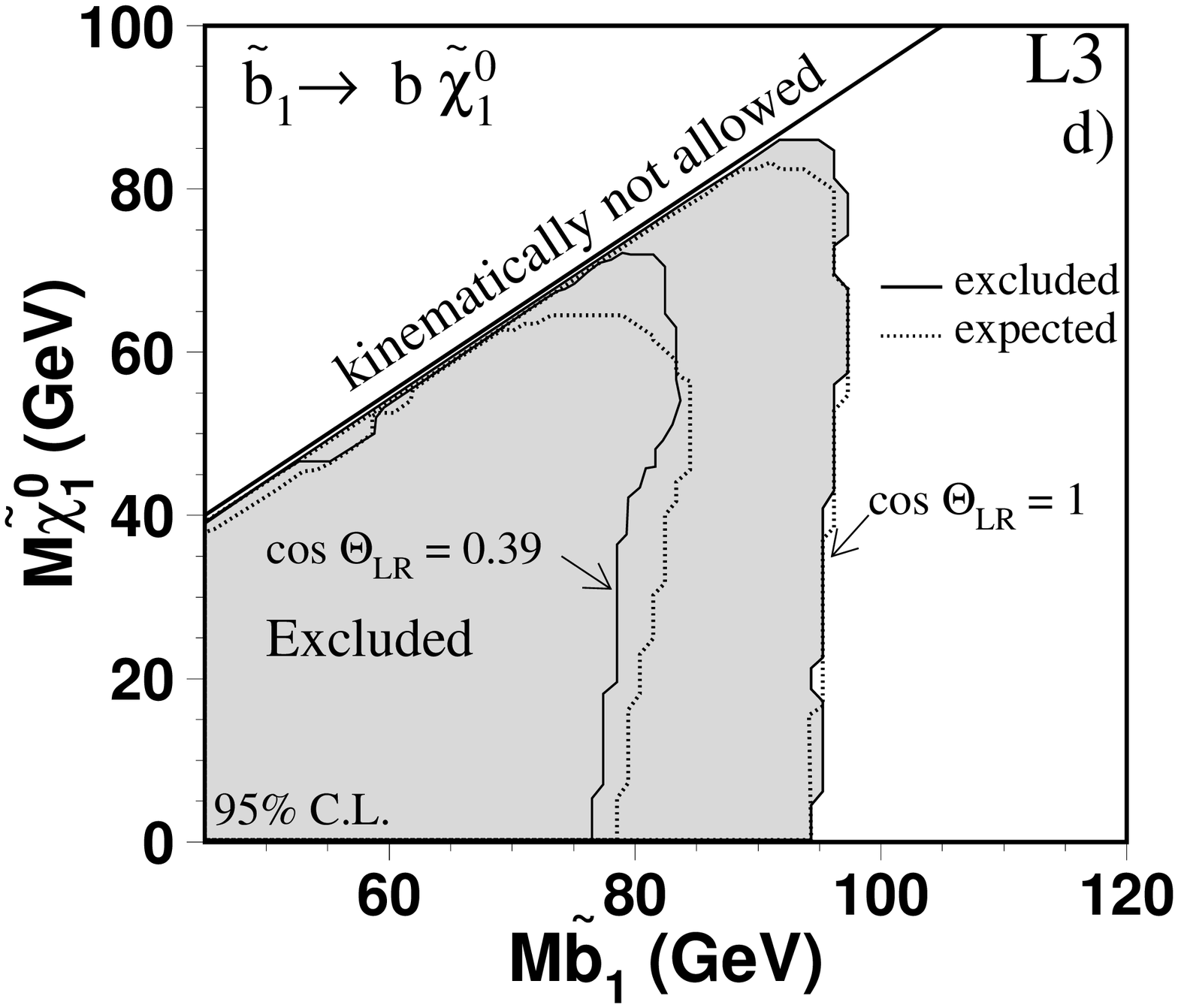}\\
    \end{tabular}
  \end{center}
\caption{\label{fig:exclusion}Regions excluded in the planes a), b) and c)
      $M_{\neutralino{1}} - M_{\susy{\rm t}_1}$ and d)
      $M_{\neutralino{1}} - M_{\susy{\rm b}_1}$.
	  The
	  MSSM decay modes: a) $\rm
	  \susy{t}_1\rightarrow c \neutralino{1}$, b)  $\rm
	  \susy{t}_1\rightarrow b \ell \susy{\nu}$, c)  $\rm
	  \susy{t}_1\rightarrow b \tau \susy{\nu}$ and d)  $\rm
	  \susy{b}_1\rightarrow b \neutralino{1}$ are studied.
	  Different values of the mixing angles are considered.}
\end{figure}

Figure \ref{fig:exclusion}a shows the excluded $\qst$ mass region
as a function of $M_{\qst}$ and $M_{\chna}$ at
$\cos\theta_{\mathrm{LR}}=1$ and $\cos\theta_{\mathrm{LR}}=0.57$ for the $\qst \to \mathrm{c}
\chna$ decay. The second value of the mixing angle corresponds to
a vanishing contribution of the Z exchange in the $s$-channel production. For
this decay mode, scalar top masses below $95\gev$ are excluded at
95\% CL
under the assumptions $\cos\theta_{\mathrm{LR}}$=1 and $\Delta
M= 15-25 \gev$. For the same values of
$\Delta M$ and in the most pessimistic scenario of
$\cos\theta_{\mathrm{LR}}=0.57$, the 95\% CL mass limit is $90\gev$. The
region in which the $\qst \to \mathrm{b} \mathrm{W} \chna$ decay is
kinematically accessible and becomes the dominant decay mode, is
indicated. This decay is not considered in this analysis.

Figure~\ref{fig:exclusion}b shows the 
scalar top mass regions which are excluded  if the dominant three-body decay
$\qst \to \mathrm{b} \ell \snu$ is
kinematically accessible. 
Equal branching fractions for the decays
into e, $\mu$ or $\tau$ are assumed and 95\% CL
mass lower limits are derived as $96
\gev$ and $93 \gev$ for $\cos\theta_{\mathrm{LR}}=1$ and  $\cos\theta_{\mathrm{LR}}=0.57$,
respectively. The corresponding exclusion limits for the scalar top decay
$\qst \to \mathrm{b} \tau \snu$ are shown in the Figure~\ref{fig:exclusion}c.
Mass lower limits at 95\% CL in the range $93-95 \gev$ are
obtained, assuming $\Delta M >15 \gev$.

Figure \ref{fig:exclusion}d shows the region excluded as a
function of  $M_{\tilde{\rm{b}}_1}$ and 
$M_{\neutralino{1}}$ considering the
$\rm\tilde{b}_1\rightarrow b\neutralino{1}$ decay
for $\cos\theta_{\mathrm{LR}}=1$ and
$\cos\theta_{\mathrm{LR}}=0.39$. The latter value corresponds to a
vanishing contribution of the Z exchange in the $s$-channel
production. Scalar bottom masses below $95 \gev$ are excluded 
at 95\% CL
assuming $\cos\theta_{\mathrm{LR}}$=1 and $\Delta M = 15-25
\gev{}$. For
$\cos\theta_{\mathrm{LR}}=0.39$, the 95\% CL mass lower limit is 81
\gev{}.

\begin{figure}
  \begin{center}
    \begin{tabular}{cc}
      \includegraphics*[width=0.4\textwidth]{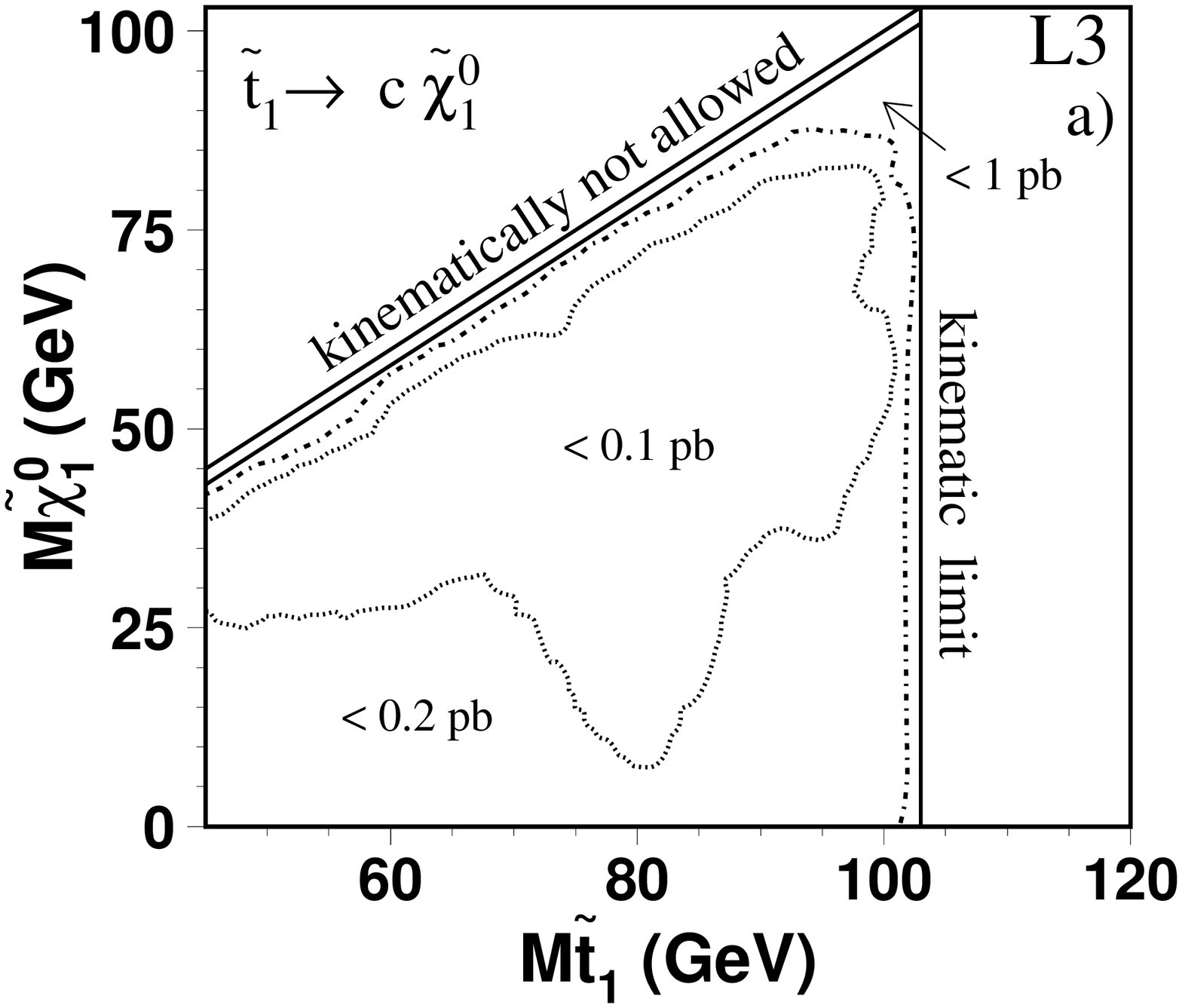}&
      \includegraphics*[width=0.4\textwidth]{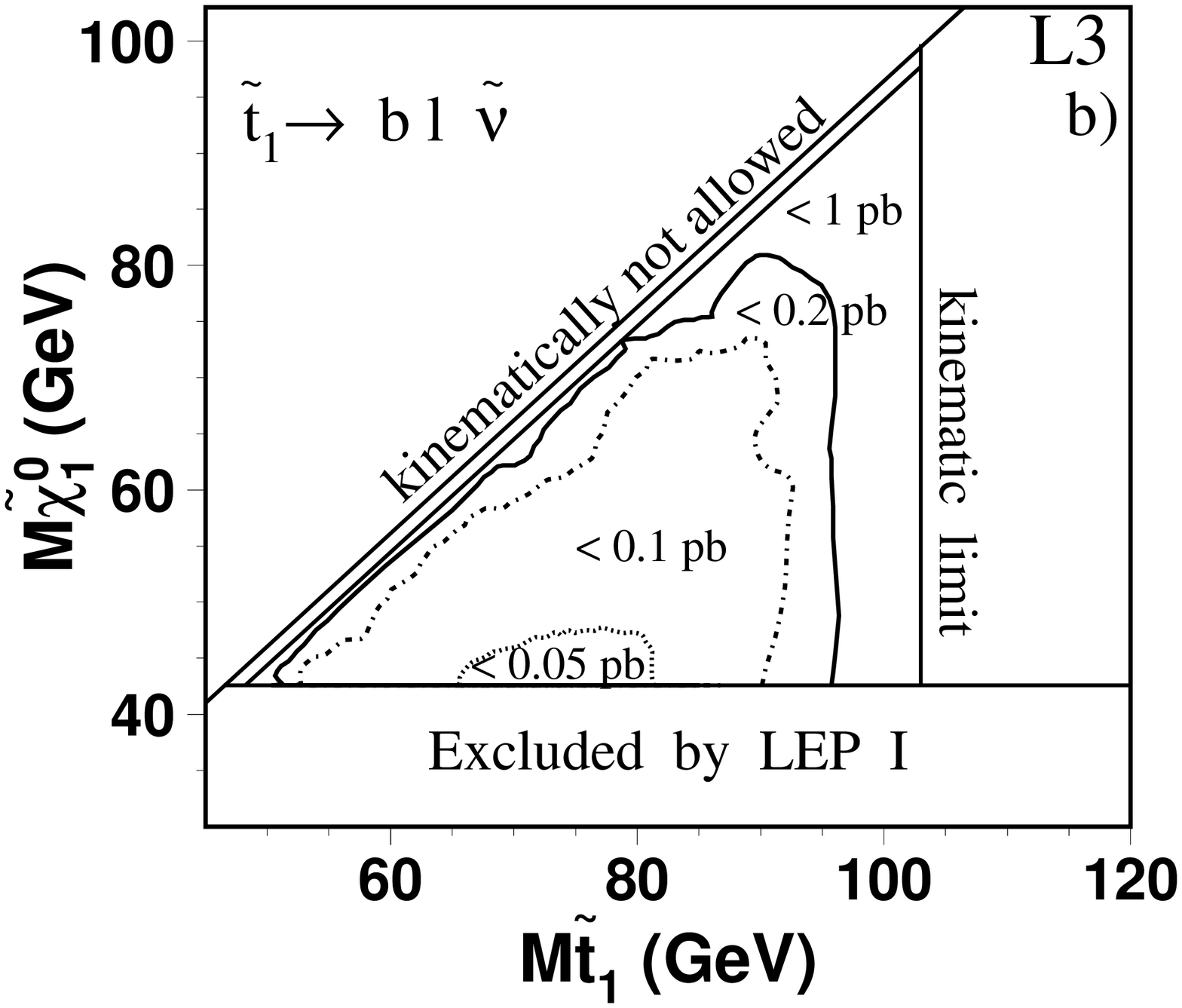}\\
      \includegraphics*[width=0.4\textwidth]{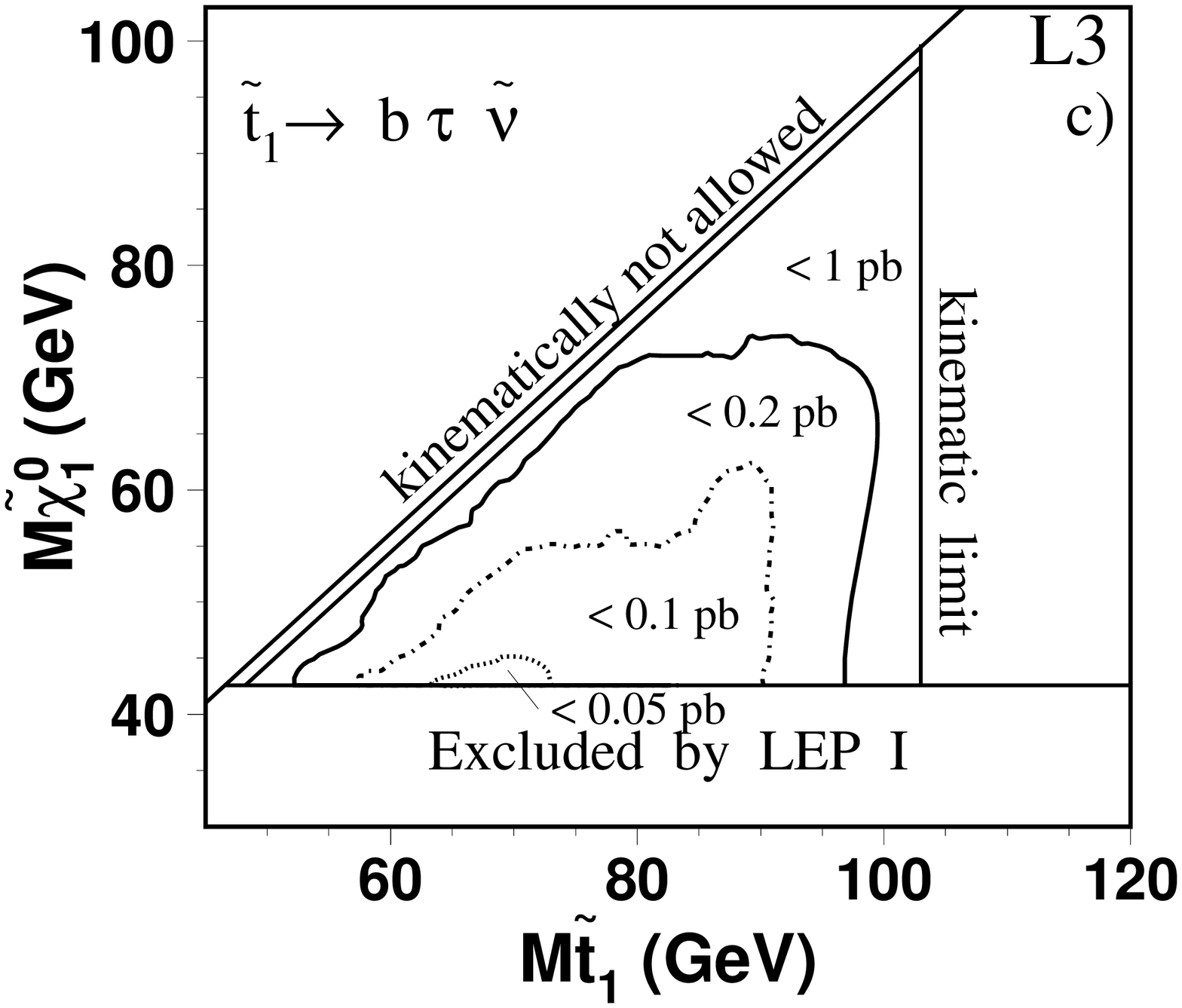}&
      \includegraphics*[width=0.4\textwidth]{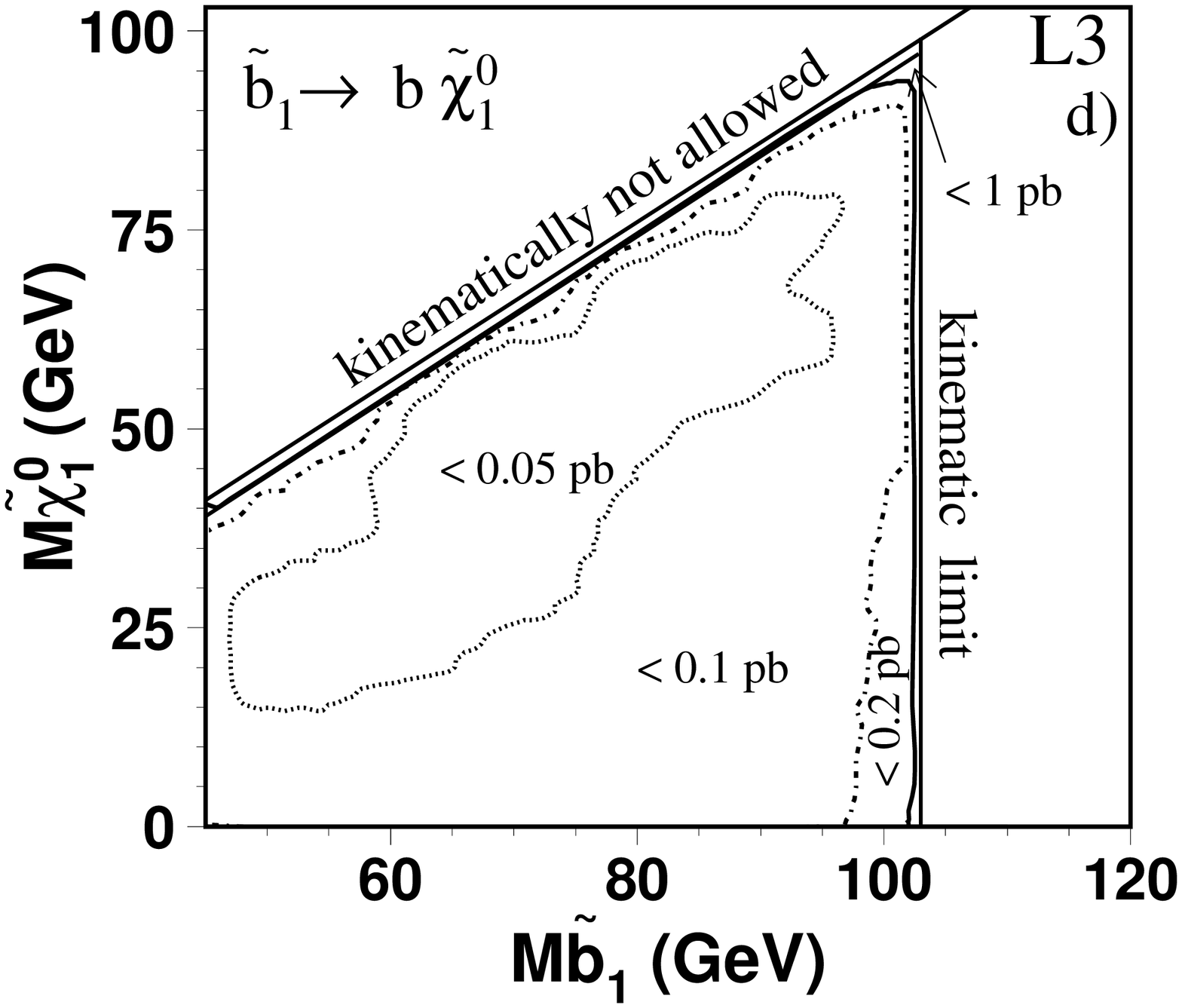}\\
    \end{tabular}
  \end{center}
      \caption{\label{fig:xbrexcl1} Model independent upper limits on the
 a), b) and c) $\mathrm{e^+e^-}
\to \qst \qast$ and d) $\mathrm{e^+e^-}
\to \qsb \qasb$ production cross sections
 multiplied by the branching ratio of the decay mode: 
a) $\rm \susy{t}_1\rightarrow c \neutralino{1}$, 
b) $\rm \susy{t}_1\rightarrow b \ell \susy{\nu}$, 
c)   $\rm \susy{t}_1\rightarrow b \tau \susy{\nu}$
and d)  $\rm \susy{b}_1\rightarrow b \neutralino{1}$.}
\end{figure}

\begin{figure}
  \begin{center}
    \begin{tabular}{c}
      \includegraphics*[width=0.55\textwidth]{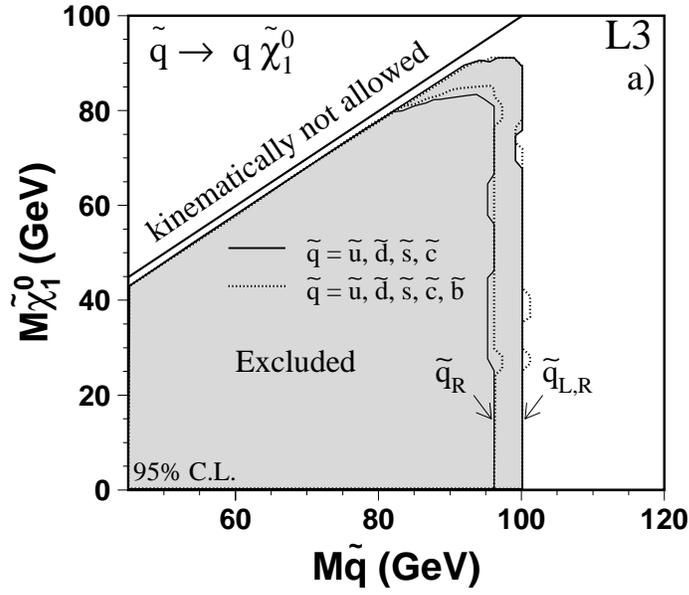}\\
      \includegraphics*[width=0.55\textwidth]{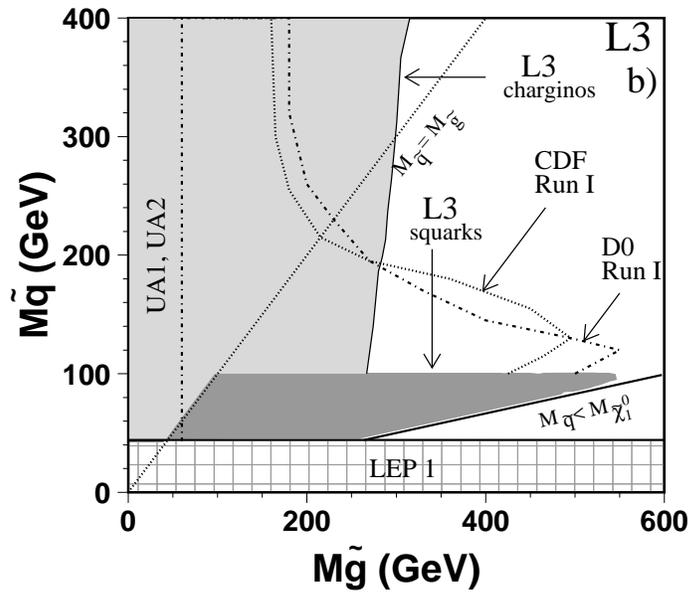}\\
    \end{tabular}
  \end{center}
\caption{\label{fig:squarks} a) MSSM exclusion limits in
the $M_{\neutralino{1}} - M_{\susy{\rm q}}$ plane
for degenerate scalar quarks decaying via
$\tilde{\mathrm{q}} \to \mathrm{q} \chna$. b) excluded
regions in the $M_{\mathrm{\tilde{g}}}-M_{\mathrm{\tilde{q}}}$ plane. The dark shaded area is excluded
by the search for scalar quarks of the first two families, assuming 
mass degeneracy among different flavours and between left-
and right-handed scalar quarks. The light shaded area illustrates indirect
limits on the gluino mass, derived from the chargino, neutralino
and scalar lepton searches. The regions excluded by the CDF and D0
collaborations~\protect\cite{CDFD0} are valid for $\tan\beta=4$
and $\mu=-400$ \gev{}. The exclusions obtained by the UA1 and
UA2~\protect\cite{ua1ua2} collaborations are also shown.}
\end{figure}

For scalar quarks of the first two generations, the same selection
efficiencies are assumed as for the $\rm\tilde{t}_1\rightarrow c\tilde{\chi}^0_1$ decay
because of the similar event topologies. The cross section limits
given in Figure~\ref{fig:xbrexcl1}a are then interpreted in terms
of degenerate scalar quark masses. Figure~\ref{fig:squarks}a
shows the scalar quark mass lower limits as a function of the \neutralino{1}
mass. Two scenarios are considered: left- and right-handed scalar quark
degeneracy or only right-handed scalar quark production. In the first
case, with four degenerate scalar quark flavours, the 95\% CL mass
limit is $99.5 \gev$ at  for  $\Delta M > 10  \gev$. In the case of only
right-handed scalar quark production, the  95\% CL mass lower limit is 97
\gev{}. 
Regions excluded in the hypotheses that all scalar quarks but the scalar top are degenerate are also
shown.

Assuming gaugino unification at the GUT scale, the results for the
four degenerate scalar quarks are reinterpreted on the plane of the
scalar
quark and gluino masses,  as shown
in Figure~\ref{fig:squarks}b. In addition, gaugino
unification~\cite{mssm1} allows a transformation of the absolute
limit on $M_2$, obtained from the chargino and neutralino~\cite{chargino} 
as well as scalar lepton searches, into a lower limit on the
gluino mass, also shown in Figure~\ref{fig:squarks}b. The ISAJET
program~\cite{isajet} is used for the calculation of the
exclusion contours. For
$\tan\beta=4$, gluino masses up to about 270$-$310 \gev{} are
excluded at 95\% CL.

In conclusion, no evidence for the production of scalar lepton
and quarks is observed in the data set collected by the L3
experiment at LEP. Stringent  upper limits on the cross sections for
the production of these scalar particles are derived, which correspond
to lower mass limits in the MSSM.

%%%%%%%%%%%%%%%%%%%%%%%%%%%%%%%%%%%%%%%%%%%%%%%%%%%%%%%%%%%%%%%%%%%%%%%%%%%%%%%
%
% Acknowledgements
%
%%%%%%%%%%%%%%%%%%%%%%%%%%%%%%%%%%%%%%%%%%%%%%%%%%%%%%%%%%%%%%%%%%%%%%%%%%%%%%%
%
%
%%%%%%%%%%%%%%%%%%%%%%%%%%%%%%%%%%%%%%%%%%%%%%%%%%%%%%%%%%%%%%%%%%%%%%%%%%%%%%%%
%                              BIBLIOGRAPHY
%%%%%%%%%%%%%%%%%%%%%%%%%%%%%%%%%%%%%%%%%%%%%%%%%%%%%%%%%%%%%%%%%%%%%%%%%%%%%%%%

\bibliographystyle{l3stylem}

%
%%%%%%%%%%%%%%%%%%%%%%%%%%%%%%%%%%%%%%%%%%%%%%%%%%%%%%%%%%%%%%%%%%%%%%%%%%%%%%
% Author List
%%%%%%%%%%%%%%%%%%%%%%%%%%%%%%%%%%%%%%%%%%%%%%%%%%%%%%%%%%%%%%%%%%%%%%%%%%%%%%
%
\newpage
\typeout{   }     
\typeout{Using author list for paper 261 -  }
\typeout{$Modified: Jul 15 2001 by smele $}
\typeout{!!!!  This should only be used with document option a4p!!!!}
\typeout{   }
%
%
%
%  L A T E X  version!!
%
%
% Make sure that the Lep package has been used!
%\input{Lep.sty}%
%
%\ifx\LepCalled\undefined%
%\typeout{     }%
%\typeout{!!!!!!!!!!!!!!!!!!!!!!!!!!!!!!!!!!!!!!!!!!!!!!!!!!!!!!!!!!!}%
%\typeout{Yikes.  You haven't used the Lep package!}%
%\typeout{Please put \protect\usepackage\protect{Lep\protect} in your preamble,
%         followed by}%
%\typeout{\protect\Lep\protect{1\protect} or \protect\Lep\protect{2\protect}}%
%\typeout{     }%
%\typeout{For now you will get a Lep phase 2 authorlist (may not be right!).}%
%\typeout{!!!!!!!!!!!!!!!!!!!!!!!!!!!!!!!!!!!!!!!!!!!!!!!!!!!!!!!!!!!}%
%\typeout{     }%
%\Lep{2}\fi%

\newcount\tutecount  \tutecount=0
\def\tutenum#1{\global\advance\tutecount by 1 \xdef#1{\the\tutecount}}
\def\tute#1{$^{#1}$}
\tutenum\aachen            % 1
\tutenum\nikhef            % 2
\tutenum\mich              % 3
\tutenum\lapp              % 4
\tutenum\basel             % 5
\tutenum\lsu               % 6
\tutenum\beijing           % 7
\tutenum\bologna           % 8
\tutenum\tata              % 9 
\tutenum\ne                % 10
\tutenum\bucharest         % 11
\tutenum\budapest          % 12
\tutenum\mit               % 13
\tutenum\panjab            % 14 
\tutenum\debrecen          % 15
\tutenum\dublin            % 16
\tutenum\florence          % 17
\tutenum\cern              % 18
\tutenum\wl                % 19
\tutenum\geneva            % 20
\tutenum\hefei             % 21
\tutenum\lausanne          % 22
\tutenum\lyon              % 23
\tutenum\madrid            % 24
\tutenum\florida           % 25
\tutenum\milan             % 26
\tutenum\moscow            % 27
\tutenum\naples            % 29
\tutenum\cyprus            % 30
\tutenum\nymegen           % 31
\tutenum\caltech           % 32
\tutenum\perugia           % 33
\tutenum\peters            % 34
\tutenum\cmu               % 35
\tutenum\potenza           % 36
\tutenum\prince            % 37
\tutenum\riverside         % 38
\tutenum\rome              % 39
\tutenum\salerno           % 40
\tutenum\ucsd              % 41
\tutenum\sofia             % 42
\tutenum\korea             % 43
\tutenum\purdue            % 44
\tutenum\psinst            % 45
\tutenum\zeuthen           % 46
\tutenum\eth               % 47
\tutenum\hamburg           % 48
\tutenum\taiwan            % 49
\tutenum\tsinghua          % 50

{
\parskip=0pt
\noindent
{\bf The L3 Collaboration:}
\ifx\selectfont\undefined%  old style font selection
 \baselineskip=10.8pt
 \baselineskip\baselinestretch\baselineskip
 \normalbaselineskip\baselineskip
 \ixpt
\else%                      new style font selection
 \fontsize{9}{10.8pt}\selectfont
\fi
\medskip
\tolerance=10000
\hbadness=5000
\raggedright
\hsize=162truemm\hoffset=0mm
\def\r{\rlap,}
\noindent

P.Achard\r\tute\geneva\ 
O.Adriani\r\tute{\florence}\ 
M.Aguilar-Benitez\r\tute\madrid\ 
J.Alcaraz\r\tute{\madrid}\ 
G.Alemanni\r\tute\lausanne\
J.Allaby\r\tute\cern\
A.Aloisio\r\tute\naples\ 
M.G.Alviggi\r\tute\naples\
H.Anderhub\r\tute\eth\ 
V.P.Andreev\r\tute{\lsu,\peters}\
% ...
\newpage
%\rule{\textwidth}{0.4pt}
\begin{list}{A}{\itemsep=0pt plus 0pt minus 0pt\parsep=0pt plus 0pt minus 0pt
                \topsep=0pt plus 0pt minus 0pt}
\item[\aachen]
 III. Physikalisches Institut, RWTH, D-52056 Aachen, Germany$^{\S}$
\item[\nikhef] National Institute for High Energy Physics, NIKHEF, 
     and University of Amsterdam, NL-1009 DB Amsterdam, The Netherlands
\item[\mich] University of Michigan, Ann Arbor, MI 48109, USA
\item[\lapp] Laboratoire d'Annecy-le-Vieux de Physique des Particules, 
     LAPP,IN2P3-CNRS, BP 110, F-74941 Annecy-le-Vieux CEDEX, France
\item[\basel] Institute of Physics, University of Basel, CH-4056 Basel,
     Switzerland
\item[\lsu] Louisiana State University, Baton Rouge, LA 70803, USA
\item[\beijing] Institute of High Energy Physics, IHEP, 
  100039 Beijing, China$^{\triangle}$ 
\item[\bologna] University of Bologna and INFN-Sezione di Bologna, 
     I-40126 Bologna, Italy
\item[\tata] Tata Institute of Fundamental Research, Mumbai (Bombay) 400 005, India
\item[\ne] Northeastern University, Boston, MA 02115, USA
\item[\bucharest] Institute of Atomic Physics and University of Bucharest,
     R-76900 Bucharest, Romania
\item[\budapest] Central Research Institute for Physics of the 
     Hungarian Academy of Sciences, H-1525 Budapest 114, Hungary$^{\ddag}$
\item[\mit] Massachusetts Institute of Technology, Cambridge, MA 02139, USA
\item[\panjab] Panjab University, Chandigarh 160 014, India.
\item[\debrecen] KLTE-ATOMKI, H-4010 Debrecen, Hungary$^\P$
\item[\dublin] Department of Experimental Physics,
  University College Dublin, Belfield, Dublin 4, Ireland
\item[\florence] INFN Sezione di Firenze and University of Florence, 
     I-50125 Florence, Italy
\item[\cern] European Laboratory for Particle Physics, CERN, 
     CH-1211 Geneva 23, Switzerland
\item[\wl] World Laboratory, FBLJA  Project, CH-1211 Geneva 23, Switzerland
\item[\geneva] University of Geneva, CH-1211 Geneva 4, Switzerland
\item[\hefei] Chinese University of Science and Technology, USTC,
      Hefei, Anhui 230 029, China$^{\triangle}$
\item[\lausanne] University of Lausanne, CH-1015 Lausanne, Switzerland
\item[\lyon] Institut de Physique Nucl\'eaire de Lyon, 
     IN2P3-CNRS,Universit\'e Claude Bernard, 
     F-69622 Villeurbanne, France
\item[\madrid] Centro de Investigaciones Energ{\'e}ticas, 
     Medioambientales y Tecnol\'ogicas, CIEMAT, E-28040 Madrid,
     Spain${\flat}$ 
\item[\florida] Florida Institute of Technology, Melbourne, FL 32901, USA
\item[\milan] INFN-Sezione di Milano, I-20133 Milan, Italy
\item[\moscow] Institute of Theoretical and Experimental Physics, ITEP, 
     Moscow, Russia
\item[\naples] INFN-Sezione di Napoli and University of Naples, 
     I-80125 Naples, Italy
\item[\cyprus] Department of Physics, University of Cyprus,
     Nicosia, Cyprus
\item[\nymegen] University of Nijmegen and NIKHEF, 
     NL-6525 ED Nijmegen, The Netherlands
\item[\caltech] California Institute of Technology, Pasadena, CA 91125, USA
\item[\perugia] INFN-Sezione di Perugia and Universit\`a Degli 
     Studi di Perugia, I-06100 Perugia, Italy   
\item[\peters] Nuclear Physics Institute, St. Petersburg, Russia
\item[\cmu] Carnegie Mellon University, Pittsburgh, PA 15213, USA
\item[\potenza] INFN-Sezione di Napoli and University of Potenza, 
     I-85100 Potenza, Italy
\item[\prince] Princeton University, Princeton, NJ 08544, USA
\item[\riverside] University of Californa, Riverside, CA 92521, USA
\item[\rome] INFN-Sezione di Roma and University of Rome, ``La Sapienza",
     I-00185 Rome, Italy
\item[\salerno] University and INFN, Salerno, I-84100 Salerno, Italy
\item[\ucsd] University of California, San Diego, CA 92093, USA
\item[\sofia] Bulgarian Academy of Sciences, Central Lab.~of 
     Mechatronics and Instrumentation, BU-1113 Sofia, Bulgaria
\item[\korea]  The Center for High Energy Physics, 
     Kyungpook National University, 702-701 Taegu, Republic of Korea
\item[\purdue] Purdue University, West Lafayette, IN 47907, USA
\item[\psinst] Paul Scherrer Institut, PSI, CH-5232 Villigen, Switzerland
\item[\zeuthen] DESY, D-15738 Zeuthen, Germany
\item[\eth] Eidgen\"ossische Technische Hochschule, ETH Z\"urich,
     CH-8093 Z\"urich, Switzerland
\item[\hamburg] University of Hamburg, D-22761 Hamburg, Germany
\item[\taiwan] National Central University, Chung-Li, Taiwan, China
\item[\tsinghua] Department of Physics, National Tsing Hua University,
      Taiwan, China
\item[\S]  Supported by the German Bundesministerium 
        f\"ur Bildung, Wissenschaft, Forschung und Technologie
\item[\ddag] Supported by the Hungarian OTKA fund under contract
numbers T019181, F023259 and T037350.
\item[\P] Also supported by the Hungarian OTKA fund under contract
  number T026178.
\item[$\flat$] Supported also by the Comisi\'on Interministerial de Ciencia y 
        Tecnolog{\'\i}a.
\item[$\sharp$] Also supported by CONICET and Universidad Nacional de La Plata,
        CC 67, 1900 La Plata, Argentina.
\item[$\triangle$] Supported by the National Natural Science
  Foundation of China.
\end{list}
}
\vfill

%%% Local Variables: 
%%% mode: latex
%%% TeX-master: t
%%% End:

\newpage

%%%%%%%%%%%%%%%%%%%%%%%%%%%%%%%%%%%%%%%%%%%%%%%%%%%%%%%%%%%%%%%%%%%%%%%%%%%%%%%%
%                              Tables
%%%%%%%%%%%%%%%%%%%%%%%%%%%%%%%%%%%%%%%%%%%%%%%%%%%%%%%%%%%%%%%%%%%%%%%%%%%%%%%%
\vspace{-10mm}
\begin{table}
\begin{center}

\begin{tabular}{|l||c|c|c||c|c|c||c|c|c|}\hline
          &
\multicolumn{3}{c||}{$\tilde{\rm e}$ } & \multicolumn{3}{c||}{$\tilde{\mu}$ } &
\multicolumn{3}{c|}{$\tilde{\tau}$} \\
  & \multicolumn{3}{c||}{ $M_{\tilde{\rm e}}=94\GeV$ } &
\multicolumn{3}{c||}{$M_{\tilde{\mu}}=90\GeV$ } &
\multicolumn{3}{c|}{$M_{\tilde{\tau}}=80\GeV$ } \\ \cline{2-10}
   &  $N_{{D}}$ & $N_{{SM}}$ & $\varepsilon$ (\%)
   &  $N_{{D}}$ & $N_{{SM}}$ & $\varepsilon$ (\%)
   &  $N_{{D}}$ & $N_{{SM}}$ & $\varepsilon$ (\%) \\ \hline
 Low   $\Delta M$  &
 79   &  84    & 10   &
 151   &  138    & 29   &
 317  &  270    & $\phantom{0}$3    \\
 Medium $\Delta M$ &
  19   &   25   & 45   &
  $\phantom{0}$46  &    $\phantom{0}$47   & 52   &
  146   &  124  & 29  \\
 High  $\Delta M$  &
  50   &   53   & 35   &
  108  &   105  & 57   &
  122 &   123   & 29  \\

\hline
\end{tabular}
\caption{\label{slepton} Results of the scalar lepton analysis:
number of observed events, $N_{{D}}$, SM background
expectations, $N_{{SM}}$, and efficiencies, $\varepsilon$, at
$\sqrt{s}=205\GeV$ for the scalar electron, muon and tau selections at
low  ($Z <  10\GeV$), medium ($10\GeV < Z
< 30 \GeV$) and high $\Delta M$ ($Z >  30 \GeV$) for different values of the 
scalar lepton masses.}
\end{center}
\end{table}

\begin{table}
\begin{center}

\begin{tabular}{|l||c|c|c||c|c|c||c|c|c||c|c|c|}\hline
\rule{0pt}{14pt} & \multicolumn{3}{c||}{$\qst \to \mathrm{c} \chna$ } &
\multicolumn{3}{c||}{$\qst \to \mathrm{b}\ell \snu$ } &
\multicolumn{3}{c||}{$\qst \to \mathrm{b}\tau \snu$ } &
\multicolumn{3}{c|}{$\qsb \to \mathrm{b} \chna$ }
\\ \cline{2-13}

  &  $N_{{D}}$ & $N_{{SM}}$  & $\varepsilon$ (\%)
  &  $N_{{D}}$ & $N_{{SM}}$  & $\varepsilon$ (\%)
  &  $N_{{D}}$ & $N_{{SM}}$  & $\varepsilon$ (\%)
  &  $N_{{D}}$ & $N_{{SM}}$ & $\varepsilon$ (\%)   \\ \hline
Very low  $\Delta M$ &
 23   &   21.6   & 18&
  2   &   2.2   & $\phantom{0}5$ &
  1   &    1.3   & $\phantom{0}6$ &
  1   &    3.8 & 13\\
Low $\Delta M$&
  $\phantom{0}$1   &    $\phantom{0}$3.1   & 22&
  0   &    0.4   & 14&
  0   &    1.6   & 16 &
  1   &    2.3  &22\\
Medium  $\Delta M$   &
  $\phantom{0}$4   &    $\phantom{0}$1.3    & 36&
  2   &    1.4    & 18&
  2   &    0.5    & 23&
  2   &    1.5   & 42\\
  High  $\Delta M$     &
  $\phantom{0}$1   &    $\phantom{0}$1.9    & 15&
  1   &    0.7    & 13&
  3   &    0.7    & 25&
  2   &    1.6  & 21\\
\hline
\end{tabular}
\caption{\label{squark} Results of the scalar quark analysis: number of
observed events, $N_{{D}}$, SM background expectations,
$N_{{SM}}$, and efficiencies, $\varepsilon$, for a 90 \gev\ scalar
quark at very low (5--10 \gev{}), low 
 (10--20 \gev{}), medium (20--40 \gev{}) and high
$\Delta M$ ($\geq$ 40 \gev{}) at $\sqrt{s}=205\GeV$.}
\end{center}
\end{table}

\begin{table}[h!]
\begin{center}
\begin{tabular}{|rcl|rcl|c|c|}
  \hline
\multicolumn{3}{|c}{Process} &\multicolumn{3}{|c|}{Decay Mode} & $N_{D}$ & $N_{SM}$\\
\hline\hline
\rule{0pt}{14pt}$  \epem $&$ \rightarrow $&$ \susy{e} \bar{\susy{e}}$ & $\susy{e} $&$\rightarrow $&$\neutralino{1}\rm e$ & 143 & $153\phantom{.}$ \\
\rule{0pt}{14pt}$  \epem $&$ \rightarrow $&$ \susy{\mu} \bar{\susy{\mu}}$ & $\susy{\mu} $&$\rightarrow $&$\neutralino{1} \mu $ &  269 & $253\phantom{.}$ \\
\rule{0pt}{14pt}$  \epem $&$ \rightarrow $&$ \susy{\tau} \bar{\susy{\tau}}$ & $\susy{\tau} $&$\rightarrow $&$\neutralino{1} \tau $ &  410 & $381\phantom{.}$\\
\rule{0pt}{14pt}$ \epem  $&$ \rightarrow $&$ \rm\susy{e}_{R} \susy{e}_{L}$ & $\susy{e}_{\rm L,R} $&$\rightarrow $&$\neutralino{1}\rm e $ & $\phantom{0}$45 & $\phantom{0}$44.6  \\
\rule{0pt}{14pt}$ \epem  $&$ \rightarrow $&$ \susy{b} \bar{\susy{b}}$ & $ \susy{b} $&$\rightarrow $&$\neutralino{1} \rm b$ & $\phantom{00}$6 & $\phantom{00}$7.7  \\
\rule{0pt}{14pt}$ \epem  $&$ \rightarrow $&$ \susy{t} \bar{\susy{t}}$& $ \susy{t} $&$\rightarrow $&$\neutralino{1} \rm c$& $\phantom{0}$29 & $\phantom{0}$26.5 \\
\rule{0pt}{14pt}$ \epem  $&$ \rightarrow $&$ \susy{t} \bar{\susy{t}}$& $\susy{t} $&$\rightarrow $&$\susy{\nu} \rm b \ell$& $\phantom{00}$4 & $\phantom{00}$4.0 \\
\rule{0pt}{14pt}$ \epem  $&$ \rightarrow $&$ \susy{t} \bar{\susy{t}}$& $\susy{t} $&$\rightarrow $&$\susy{\nu} \rm b \tau $& $\phantom{00}$5 & $\phantom{00}$3.9 \\

  \hline
\end{tabular}
\caption{Summary of the number of observed data events, $N_{D}$, and SM
background expectations, $N_{SM}$, for all the studied
topologies\label{results}}
\end{center}
\end{table}
%%%%%%%%%%%%%%%%%%%%%%%%%%%%%%%%%%%%%%%%%%%%%%%%%%
%                              Figures
%%%%%%%%%%%%%%%%%%%%%%%%%%%%%%%%%%%%%%%%%%%%%%%%%%%%%%%%%%%%%%%%%%%%%%%%%%%%%%%%

\clearpage

\end{document}

%%% Local Variables:
%%% mode: latex
%%% TeX-master: t
%%% End: